\documentclass[onecolumn,amsmath,amssymb,nofootinbib,12pt]{article}
 
 \ifx\pdfoutput\not\undefined
 \pdfoutput=1
\fi
\usepackage{jheppub}
\usepackage{ifpdf}
\usepackage{mdframed}
\usepackage{comment}
\usepackage{graphicx,subcaption}
\graphicspath{ {figures/} }
\usepackage{epsfig}
\usepackage{pdfpages}
\usepackage{bbm}
\usepackage{graphicx,epstopdf}
\usepackage[numbers,sort&compress]{natbib}
\usepackage[makeroom]{cancel}
\usepackage{hyperref}
\hypersetup{pdftex,colorlinks=true,allcolors=blue}
\usepackage{array}
\usepackage[export]{adjustbox}
\usepackage{amsthm}
\usepackage[normalem]{ulem}

\theoremstyle{definition}

\usepackage[utf8]{inputenc}

\usepackage{tocloft}
\usepackage{bbold}
\usepackage{tikz}
\usetikzlibrary{backgrounds}
\usetikzlibrary{patterns}
\usetikzlibrary{arrows.meta}
\usetikzlibrary{shapes,snakes}
\tikzstyle{bag} = [align=center]
\usetikzlibrary{decorations.pathmorphing}
\def\bea{\begin{eqnarray}}
\def\eea{\end{eqnarray}}

\usepackage{hyperref}
\usepackage{subcaption}

\newcommand{\ie}{{\it i.e.,}\ }

 \newcommand{\badat}{\begin{alignedat}}
 \newcommand{\eadat}{\end{alignedat}}
 
\def\ee{\end{equation}}

\def\p{\partial}

\usepackage{xcolor}

\newcommand{\pink}[1]{\textcolor{\pink}{#1}}

\definecolor{dblue}{rgb}{0.2,0.50,0.80}

\newcommand{\al}{\alpha}
\newcommand{\be}{\beta}
\newcommand{\del}{\delta}
\newcommand{\bep}{\bar{\varepsilon}}
\newcommand{\vep}{\varepsilon}
\newcommand{\bpsi}{\bar{\psi}}

\newcommand{\ga}{\gamma}
\newcommand{\dga}{\dot{\gamma}}
\newcommand{\dbe}{\dot{\beta}}
\newcommand{\dal}{\dot{\alpha}}

\newcommand{\bp}{\overline{\partial}}
\newcommand{\N}{\mathcal{N}}

\newcommand{\ov}[1]{\overline{#1}}
\newcommand{\ep}{\epsilon}
\newcommand{\epb}{\overline{\epsilon}}
\newcommand{\PT}{\mathbb{PT}}
\newcommand{\CP}{\mathbb{CP}}

\def\N{\mathcal{N}}
\def\C{\mathcal{C}}
\def\O{\mathcal{O}}
\def\A{\mathcal{A}}

\def\H{\mathcal{H}}

\def\F{\mathcal{F}}

\def\bz{{\bar z}}

\def\bchi{{\bar{\chi}}}

\definecolor{vert}{rgb}{0.1367 0.543 0.1367}

\title{Supersymmetric twists in twistor space and holography
 }

\author[a]{Matheus Balisa,}
\author[a]{and Eduardo Casali}

\affiliation[a]{Departamento de Física Matemática, Universidade de São Paulo, São Paulo, Brasil}

\emailAdd{matheusbalisa@usp.br}
\emailAdd{ecasali@usp.br}

\date{\today}

\abstract{We compute some supersymmetric twists of field theories in twistor space, including the minimal supersymmetric and the chiral algebra twists of supersymmetric self-dual Yang-Mills, and the minimal twist of $\mathcal{N}=1$ self-dual supergravity. In the case of $\mathcal{N}=4$ we also find their holographic duals in the framework of chiral holography. We find that the minimal twist of gauge theories in twistor space localizes them to spacetime, making the choice of complex structure manifest, and reproducing the minimal twist on spacetime. 
For superconformal theories we apply a further twist which localizes the theory to a plane contained on spacetime, reproducing the chiral algebra twist of $\mathcal{N}=4$ sYM. We show that the bulk duals of these twists also localize reproducing the results from twisted holography.}
\begin{document}

\maketitle

\section{Introduction}

Supersymmetric field theories have sectors preserving various amounts of supersymmetry which are usually much simpler to study than the full theory. These are called BPS sectors and can be characterized as cohomology classes of one, or a sum of, supercharges. These supersymmetric BPS sectors can be studied by a \textit{twist} of the original theory \cite{Witten:1988ze}, the name coming from the original construction where part of the R-symmetry is promoted to a spacetime symmetry, thus twisting the spin of the fields by R-charges. A second step is to promote a part of the global supersymmetry to a gauge redundancy, such that the twisted theory can be identified as a gauge-fixing of a more fundamental topological-holomorphic theory \cite{Witten:1988ze,Johansen:1994aw,Losev:1996up,Closset:2013vra}. This addition of supercharges to the BRST operator can be done very elegantly via the BV-BRST formalism \cite{Baulieu:1990uv,Elliott:2020ecf} 
which is well-suited to more rigorous mathematical formulations of the supersymmetric twist \cite{costello2013notessupersymmetricholomorphicfield,Costello:2013zra}.

In Euclidean four dimensions, the minimal twist is described by holomorphic theories \cite{Elliott:2020ecf}. The appearance of a complex structure can be seen as a result of the choice of a supercharge to twist the theory which is parametrized by a projective Weyl spinor since the cohomology doesn't care about an overall rescaling \cite{Eager:2018dsx,Saberi:2021weg}. Taking a bilinear of this projective spinor defines a two-form that can be taken to be the top-holomorphic on spacetime seen as $\mathbb{C}^2$, thus defining a complex structure on the tangent space through its kernel. The total space of projective chiral spinors over spacetime, that is, spacetime together with all its choices of complex structure, is also know as twistor space. It is then quite natural to study the twist of supersymmetric theories in twistor space.

Twistor space carries a natural complex structure, where holomorphic field theories correspond to self-dual theories on spacetime. The supersymmetric twist of these self-dual theories coincides perturbatively with the twist of the usual theories, so for the purposes of studying BPS sectors it is enough to study their supersymmetric self-dual versions. A non-local vertex in twistor space can be added to action of supersymmetric self-dual Yang-Mills  \cite{boels_supersymmetric_2007}, recovering, perturbatively, the theories in full. This has recently been interpreted as the contributions from $D9$-branes in the bulk dual in the framework of chiral holography \cite{Sharma:2025ntb}.


The minimal twist of $\mathcal{N}=1$ super Yang-Mills in twistor space was given in  \cite{costello2013notessupersymmetricholomorphicfield} in a somewhat abstract way. Here we give a more concrete construction based on the BV-BRST methods of \cite{Baulieu:1990uv} adapted to the supersymmetric Lagrangians in twistor space as given in \cite{Boels:2006ir,Mason:2007ct}, and expand it to theories with extended supersymmetry, as well as study the equivalent of the chiral algebra twist \cite{Beem:2013sza}. We find that the minimal twist localizes the theory to a point on the twistor sphere, the fiber of the projective spinor bundle. Away from this point the action is a BRST-exact term, that is, topological while on top of this point the theory is a holomorphic field theory on $\mathbb{C}^2$ with complex structure given by this point on the twistor sphere. A subtlety that appears in twistor space, and is absent in spacetime, is that the twisting SUSY charge has non-trivial commutators with the extra gauge redundancy present in twistor space. That requires some care when formulating the BV-BRST action and reduction to spacetime.

We also construct a holographic duals to the minimal and chiral algebra twists of $\mathcal{N}=4$ SYM in the context of the recently introduced chiral holography \cite{Sharma:2025ntb}, showing that it reproduces, perturbatively, the duals conjectured in \cite{Costello:2018zrm}. In the case of the chiral algebra twist \cite{Beem:2013sza} and its dual, we find at first glance the same results as \cite{Jarov:2025qhz}, but the latter uses a fermionic symmetry which is not a supersymmetry from the point of view of spacetime, getting a chiral algebra over the twistor sphere. Our construction gives the usual localization of the chiral algebra to a plane on spacetime, which has a nice geometric interpretation in twistor space. Its dual follows from a similar construction as in \cite{Costello:2018zrm} adapted to the chiral holography setting.

\section{Supersymmetric gauge theory in twistor space}

In this section we will review the necessary aspects of supersymmetric gauge theories in twistor space that we'll need in the next sections. We'll follow the more modern presentation of \cite{Witten:2003nn,Boels:2006ir,Wolf:2010av,Adamo:2013cra}, a quick and useful introduction to twistor theory can be found in \cite{Adamo:2017qyl}.
\subsection{Supertwistor space}

Take $Z^A = (\lambda_\al,\mu^{\dal})$, with $\alpha=\{0,1\}$ and $\dal=\{1,2\}$, as homogeneous holomorphic coordinates on $\mathbb{CP}^3$. That is, $Z^A\in\mathbb{C}^4$ are taken up to a complex non-zero rescaling $Z^A\sim r Z^A$ which $r\in \mathbb{C}^*$. Holomorphic objects in $\mathbb{CP}^3$ are naturally graded under this scaling. Twistor space, denoted $\mathbb{PT}$, is obtained by removing a line from $\mathbb{CP}^3$, that is $\mathbb{PT}=\mathbb{CP}^3\backslash\{\lambda_\al=0\}$. As a manifold $\mathbb{PT}\simeq \mathbb{R}^4\times S^2$ where the first factor can be identified with Euclidean flat spacetime, and the removed line corresponds to the $S^1$ at infinity from the point of view of spacetime. As a complex manifold, twistor space can also be viewed as the total space of the holomorphic fibration $\mathcal{O}(1)^{\oplus2}\rightarrow\mathbb{CP}^1$, where $\mu^{\dal}$ are the fiber coordinates and $\lambda_\al$ are homogeneous coordinates in $\mathbb{CP}^1 $. The correspondence with spacetime is realized through the incident relations
\begin{equation}
    \mu^{\dal} = \lambda_\al x^{\al\dal}.
\end{equation}
and a choice of involution on $\PT$ which fixes a reality structure for the spacetime coordinates $x^{\al\dal}$, and a signature for the spacetime metric. 

$\N=4$ supertwistor space is obtained by adjoining to $\CP^3$ four fermionic directions with coordinates $\eta_i$, where $i=1,2,3,4$, of weight one under the complex rescaling, that is $\eta_i\in\Pi\O(1)$, where $\Pi$ denotes parity reversal. This fermionic extension of the complex projective space is denoted $\CP^{3|4}$. Removing a superline from it $\mathbb{PT}^{3|4} = \mathbb{CP}^{3|4}\backslash\mathbb{CP}^{1|4}$ 
gives the $\N=4$ supertwistor space. The homogeneous coordiantes $[Z_A,\eta_i]$ are defined up to a complex rescaling $(Z_A,\eta_i)\sim (rZ_A,r\eta_i)$ and we again parametrized the line removed by $\lambda_\al=0$. Supertwistor space can be seen as the total space of the $\mathcal{O}(1)^{\oplus2}\oplus\Pi\mathcal{O}(1)^{\oplus 4}\rightarrow \mathbb{CP}^1$ holomorphic fibration. Altough the top holomorphic form on $\PT$
\begin{equation}
D^3Z=\vep_{ABCD}Z^AdZ^BdZ^CdZ^D.
\end{equation}
is twisted by $\O(4)$ and so $\PT$ is not Calabi-Yau, supertwistor space has a canonical global holomorphic volume form 
\begin{equation}
   \Omega =D^3Zd^4\eta, \quad d^4\eta=\vep^{ijkl}d\eta_id\eta_jd\eta_kd\eta_l. 
\end{equation}
since be fermionic statistics $d\eta_i$ are twisted by $\O(-1)$. This makes $\mathbb{PT}^{3|4}$ into a super Calabi-Yau manifold.

We'll work in Euclidean signature throughout this paper which is defined by the involution
\begin{equation}
    \lambda_{\al} = (\lambda_1,\lambda_2)\rightarrow \hat{\lambda}_\al =(-\overline{\lambda_1},\overline{\lambda_2}).
\end{equation}
This obeys $\hat{\hat{\lambda}}=-\lambda$, implying that there are no non-vanishing real spinors in Euclidean signature. Using the incidence relations we can naturally define a basis of anti-holomorphic one forms \cite{Woodhouse:1985id}
\begin{equation}\label{OneFormsBasis}
    \overline{e}_0 = \frac{\hat{\lambda}^\al d\hat{\lambda}_\al}{\langle\lambda\hat \lambda\rangle^2}\quad \overline{e}^{\dal} = \frac{\hat \lambda_\al dx^{\dal\al}}{\langle\lambda\hat{\lambda}\rangle},
\end{equation}
and associated dual anti-holomorphic vectors
\begin{equation}\label{ExteriorDerivativeBasis}
    \bp_0 = \langle\lambda\hat{\lambda}\rangle{\lambda_\al}\frac{\p}{\p\hat{\lambda}_\al}\quad \bp_{\dal} ={\lambda}_\al\frac{\p}{\p x^{\al \dal}}.
\end{equation}
The holomorphic volume form becomes:
\begin{equation}
    D^3Z = \langle\lambda d\lambda\rangle d\mu^{\dal}d\mu_{\dal} =\langle\lambda d\lambda\rangle \lambda_\al\lambda_\be dx^{\dal\al}dx^\be_{\dal}.
\end{equation}

\subsection{The supersymmetry algebra}

Since we'll be making use of the supersymmetry in twistor space throughout the paper, we quickly review the realization of the $\mathfrak{psl}(4|4)$ superconformal algebra on twistor space.

It is useful to decompose the Lorentz generators into SD and ASD components $L^{\al}_\be$ and $L^{\dal}_{\dbe}$. We will allow for a central element $Z$ in the algebra $\mathfrak{psl}(4|4)$ to facilitate computations. Since the central element vanishes when working projectively \cite{Wolf:2010av} (this can be easily seen from the generator below), it will not matter in the end. One realization for this algebra is given by the following generators \cite{Wolf:2010av}:

\begin{itemize}
    \item 31 Bosonic generators: these are the conformal generators plus the R-symmetry generators and the central charge
    \begin{align*}
        &P_{\al\dal} = \lambda_\al \frac{\p}{\p\mu^{\dal}} \quad K^{\al\dal} = \mu^{\dal}\frac{\p}{\p \lambda_\al} \quad L^\al_\be = i\Big(\lambda^\be\frac{\p}{\p\lambda_\al} - \frac{1}{2}\del^\al_\be \lambda_\ga \frac{\p}{\p \lambda_\ga}\Big) \\& L^{\dal}_{\dbe} = i\Big(\mu^{\dal}\frac{\p}{\p\mu^{\dbe}} - \frac{1}{2}\del^{\dal}_{\dbe}\mu^{\dga}\frac{\p}{\p\mu^{\dga}}\Big) \quad D = -\frac{i}{2} \Big(\mu^{\dal}\frac{\p}{\p\mu^{\dal}} - \lambda_\al\frac{\p}{\p\lambda_\al}\Big)\\&Z =-\frac{i}{2} \Big(\mu^{\dal}\frac{\p}{\p\mu^{\dal}} + \lambda_\al\frac{\p}{\p\lambda_\al}+\eta_j\frac{\p}{\p \eta_j}\Big)  \quad R^i_j = -\frac{i}{2}\Big(\eta_j\frac{\p}{\p\eta_i} - \frac{1}{4}\eta_k\frac{\p}{\p\eta_k}\Big)
    \end{align*}
    \item 32 Fermionic generators: These are ordinary supercharges $Q$ and superconformal supercharges $S$:
    \begin{equation}\label{FirstBasis}
        Q^i_{\al} = \lambda_\al\frac{\p}{\p\eta_i} \quad \tilde Q_{i\dal} =  \eta_i\frac{\p}{\p\mu^{\dal} }\quad \tilde S^{\dal i} = \mu^{\dal}\frac{\p}{\p\eta_i} \quad S_i^\al = \eta_i \frac{\p}{\p\lambda_\al}
    \end{equation}
The central element can be seen to enter the algebra through the anticommutators of $Q$ and $S$:
    \begin{align}
        &\{Q_{\al}^i,S_{j}^\be\} = -i\Big(2\del^\be_\al R_j^i + \del^i_j L_\al^\be + \frac{1}{2}\del_j^i\del_\al^\be (D+Z)\Big)\label{ComutatorQandS}\\& \{\tilde Q_{i\dal},\tilde S^{j\dbe}\} = i\Big(-2\del^{\dbe}_{\dal} R_j^i + \del^i_j L_{\dal}^{\dbe} + \frac{1}{2}\del_j^i\del_{\dal}^{\dbe} (D-Z)\Big)
    \end{align}
\end{itemize}

\subsection{Self-dual supersymmetric Yang-Mills}\label{sec:SDSYM}

Witten \cite{Witten:2003nn} showed that the self-dual sector of $\N=4$ super Yang-Mills appears as the string field theory for the open sector of the B-model with target $\PT^{3|4}$. In Witten's construction this appears as the worldvolume theory of space-filling $D5$ branes wrapping $\mathbb{PT}^{3|4}$. More recently, in \cite{Sharma:2025ntb} Sharma and Skinner considered the B-model with target the total space of the fibration $X=\mathcal{O}(-1)^{\oplus 4}\rightarrow \mathbb{PT}$ wrapping a brane over the zero section $\PT$. In either case, the theory on the brane is described by holomorphic Chern-Simons on $\mathbb{PT}^{3|4}$:

\begin{equation}\label{HoloCSTwistor}
    S = \int_{\mathbb{PT}} D^3Zd^4\eta \Big(\frac{1}{2}\mathcal{A}\bp\mathcal{A} + \frac{1}{3}\A^3\Big).
\end{equation}
Where the physical fields are contained in the superconnection $\A\in\Omega^{0,1}(\PT^{3|4},\mathfrak{gl}(N,\mathbb{C}))$. The full BV complex is obtained by allowing $\mathcal{A}$ to take any anti-holomorphic form degree (\ie $\A\in \Omega^{0,*}(\mathbb{PT}^{3|4},\mathfrak{gl}(N,\mathbb{C}))$ ). The other fields in the BV multiplet are the ghosts and anti-fields. 

The component action is found by expanding the physical fields in powers of $\eta$ as:
\begin{equation}
    \A(Z,\bar Z,\eta) = a + \psi^i\eta_i + \frac{1}{2}\phi^{ij}\eta_i\eta_j + \frac{1}{3!}\chi_i\eta^{ i} +b\eta^4. 
\end{equation}
where we used the notation $\eta^{i} = \epsilon^{ijkl}\eta_j\eta_k\eta_l$, $\eta^{4} = \epsilon^{ijkl}\eta_i\eta_j\eta_k\eta_l$. All fields in the expansion are (0,1) forms with coefficients that are functions of $Z,\overline{Z}$. Since $\A$ has no projective weight, the fields $a,\psi,\phi,\chi,b$ have weights $0,-1,-2,-3,-4$ respectively. After integrating the fermionic coordinates we obtain 

\begin{equation}\label{n=4action}
    S = \int_{\mathbb{PT}}D^3Z\Big( b\ov{D}a + \psi^i\ov{D }\chi_i +\frac{1}{4} \epsilon_{ijkl}\phi^{ij}\ov{D}\phi^{kl} +  \frac{1}{2}  \ep_{ijkl}  \psi^i\psi^j\phi^{kl}\Big).
\end{equation}
The reduction to spacetime is a standard procedure that can be found in \cite{Sharma:2025ntb,Boels:2006ir} and gives $\N=4$ self-dual Yang-Mills.

To obtain theories with less supersymmetry we need to single out some fermionic directions \cite{Boels:2006ir}. For example, to obtain $\N=2$ we impose that the fields can only depend on $\eta_3$ and $\eta_4$ through the combination $\eta_3\eta_4$. Hence, we can rewrite $\A$ as 
\begin{equation}\label{N=2FieldsVector}
    \A = (a+ \psi^i\eta_i + \frac{1}{2}\epsilon^{ij}\phi\eta_i\eta_j) + \eta_3\eta_4\Big(\ov{\phi}+\ep^{ij}\chi_j\eta_i+\frac{1}{2}\epsilon^{ij}\eta_i\eta_j b\Big) = \A_2 + \eta_3\eta_4\mathcal{B}_2,
\end{equation}
where we have identified the terms inside the parenthesis with the superfields $\A_2$  and $\mathcal{B}_2$. Performing the integral over $\eta_3\eta_4$ we obtain a holomorphic $BF$ action 
\begin{equation}\label{N=2HoloBF}
    S = \int_{\mathbb{PT}} D^3Zd^2\eta\, \mathcal{B}_2 \F_2,
\end{equation}
where $\F_2 = \bp \A_2 + [\A_2,\A_2]$ is the field strength generated by $\A_2$. After further fermionic integration, the action becomes
\begin{equation}\label{N=2Action}
    S = \int_{\mathbb{PT}}D^3Z\left( b\ov{D}a + \psi^i \ov{D}\chi_i +  \ov{\phi}\,\ov{D}\phi +\frac{1}{2}\ep_{ij} \psi^i\wedge\psi^j \wedge \ov{\phi}\right)
\end{equation}
which corresponds to pure $\N=2$ self-dual Yang-Mills. To add massless matter hypermultiplets to this action we use the fact that $\N=4$ vector multiplet is given by one $\N=2$ vector multiplet and one hypermultiplet in the adjoint representation (with its CPT conjugate).  It is easy to see that the hypermultiplet should look like \cite{Ferber:1977qx,Boels:2006ir}:
\begin{equation}\label{M=2_SU[erfields]}
    \H = \rho + h^i\eta_i + \frac{1}{2}\mu\epsilon^{ij}\eta_i\eta_j \quad \tilde \H = \tilde \mu + \tilde h^i \eta_i \tilde +\frac{1}{2}\tilde \rho\epsilon^{ij}\eta_i\eta_j,
\end{equation}
with $\H$ and $\tilde\H$ fermionic, with projective weight $-1$. This is valid for any representation of the gauge group; however, in this paper we'll consider matter fields in the adjoint for simplicity. The action for the matter fields is simply
\begin{equation}\label{N=2_Matter_Action}
    S_{matter}=\int_{\mathbb{PT}} D^3Z d^2\eta\,\H\ov{D}_2\H 
\end{equation}
where $\ov{D}_2$ is the covariant derivative with respect to $\A_2$.

For $\N=1$ we follow the same procedure. We impose that the fields depend only on the combination $\eta_2\eta_3\eta_4$. Labeling $\eta_1 = \eta$, we find
\begin{equation}\label{N=1FIelds}
    \A = (a + \psi \eta) + \eta_2\eta_3\eta_4(\chi+b\eta) = \A_1 + \eta_2\eta_3\eta_4\mathcal{B}_1.
\end{equation}
After eliminating $\eta_2\eta_3\eta_4$ the action becomes a holomorphic $BF$ theory. After fermionic integration, we find 
\begin{equation}\label{N=1_Action}
    S = \int_{\mathbb{PT}} D^3Z\left( b\ov{D}a + \psi\ov{D}\chi\right).
\end{equation}
The matter fields of $\N=1$ are the chiral multiplets. The $\N=4$ vector multiplet can be described by one vector multiplet of $\N=1$ and three chiral multiplets. Hence we must have 
\begin{equation}\label{N=1matterFields}
    \C = \nu+m\eta\quad\tilde \C = \tilde m + \tilde \nu \eta,
\end{equation}
with $\C$ fermionic of weight $-1$ and $\tilde C$ is bosonic of weight $-2$. The action is 

\begin{equation}\label{N=1ChiralMultiplet}
    S = \int_{\mathbb{PT}} D^3Z\, \C\ov{D}_1\tilde \C,
\end{equation}
where $\ov{D}_1$ is the covariant derivative with respect to $\A_1$.

\section{Review of supersymmetric twists}

Before performing the supersymmetric twist in twistor space it is worthwhile to briefly review how the twist is performed on spacetime. We start with the definition of the twist following \cite{Costello:2013zra,Elliott:2020ecf,Bomans:2025klo}, implemented using the BV formalism following \cite{Baulieu:1990uv,Baulieu:2004pv,Baulieu:2010ch}, which we will then adapt to twistor space.

\subsection{The holomorphic twist on spacetime}

By supersymmetric twist we take the rater general procedure of
\begin{itemize}
    \item Picking a nilpotent supercharge, or a nilpotent linear combination of them\footnote{The space of all possible twists is called the nilpotence variety and was studied in generality in \cite{Eager:2018dsx,Garner:2022its}}.
    \item Adding the chosen supercharge to the BRST operator. The cohomology of the new BRST operator defines the twisted theory.  
\end{itemize} 


It is also common in the literature, but not necessary, to perform two additional steps:
\begin{itemize}
    \item Find a homomorphism $\phi$ that defines the action of a "new Lorentz group" on which the nilpotent supercharge transforms as a scalar. This can be achived by mixing the isometry group from spacetime with the $R$-symmetry group of the theory \ie $\phi:K' \rightarrow \text{spin}(d)\times G_R$, where $K'$ is the new Lorentz group, and $G_R$ is the R-symmetry group. 
    The spin of the fields from the twisted theory is defined accordingly to the action of $K'$.
    \item Choose a $U(1)$ subgroup of the original $R$-symmetry and spacetime symmetry groups under which the nilpotent supercharge has charge one. This $U(1)$ will define the cohomological grading (or ghost number) of the twisted theory. 
\end{itemize}

There are cases where one, or both of these steps are not possible to perform, for example, in the case of $\N=1$ we have $G_R=U(1)$ and a choice of supercharge (say $Q_\al$) breaks the spacetime symmetry to an $SU(2)\subset \text{Spin}(4)$. In $D=10$ there is not even an R-symmetry, nonetheless a notion of supersymmetric twist persists \cite{Baulieu:2010ch}.  


Performing the twist then promotes part of the global supersymmetry, generated by some supercharge $Q$, to a gauge redundancy and associated nilpotent operator $\delta_Q$. If the theory already has a BRST operator, say $\delta$ from other gauge redundancies, then we must ensure that the the new BRST operator $\delta+\delta_Q$ is nilpotent. It might be that nilpotency is obstructed by the equations of motion or that the resulting gauge algebra is reducible. In the case of twistor space we will find that the supersymmetry transformations do not commute with the BRST operator of the untwisted theory. In any case, the BV formalism can be used to define the appropriate twisted complex in an elegant way.

\subsubsection{$\N=1$}

The R-symmetry of $\N=1$ theories in four dimensions doesn't have an $SU(2)$ factor, nevertheless a version of the twist can still be done\footnote{At least perturbatively. Instanton effects might break the R-symmetry to a finite subgroup as in pure $\mathcal{N}=1$ sYM}. Full Lorentz covariance is lost but a subgroup of it still remains and we can interpret the theory as living on a space of reduced holonomy group. Concretely, the original theory has a susy transformation of the form $\del \Phi = (\vep^{\al}Q_\al +\bep^{\dal}\tilde Q_{\dal})\Phi$. Picking a particular supercharge to perform the twist is akin to promoting one of the supersymmetry parameters, say $\vep^{\al}$, to a constant background field\footnote{It can be seen as coupling to a SUGRA background where the ghost for supersymmetries has a constant vev.}. The new spacetime symmetry group is then the Lorentz transformations that leave this background field invariant. Since $\vep^{\al}$ is a Weyl spinor, and trivially pure in $D=4$, it defines a complex structure on $\mathbb{R}^4$. We expect that the spacetime symmetry group of the twisted theory preserves the choice of complex structure, that is, the twisted theory should be a holomorphic field theory on $\mathbb{C}^2$.


Lets see concretely how this works. The classical action of pure $\N=1$ sYM can be written in the Chalmers-Siegel \cite{Chalmers:1996rq} form 
\begin{equation}\label{CS_cl}
    S_{cl} = \int d^4x\left( B_{\al\be}F^{\al\be} +\frac{g^2}{2} B^2 + \psi^{\dal} D_{\dal\al}\bpsi^{\al}\right)
\end{equation}
which is perturbatively equivalent to the usual Yang-Mills action. Its supersymmetry transformations are 
\begin{align}
    &\del A_{\al\dal} = \vep_{\al}\psi_{\dal}-g^2\bep_{\dal}\ov{\psi}_\al
    \quad\del \bpsi^{\al} = \vep_{\be}B^{\al\be}\\&\del B_{\be\al} = \bep^{\dal}\p_{\dal(\al}\bpsi_{\be)}
    \quad\del \psi^{\dal} = \bep^{\dbe} \p_{\be\dbe}A^{\dal\be}-\frac{1}{2}\bep^{\dal}\p_{\ga\dga}A^{\ga\dga}.
\end{align}
The full BV action in the more general case where both supercharges are added to the BRST  was written down in \cite{Baulieu:1990uv} and here we adapt it to the Chalmers-Siegel forms of the action. It splits into three parts
\begin{equation}
    S_{BV} = S_{\text{cl}}+S_{\text{gauge}} +S_{\text{susy}},
\end{equation}
where where $S_{cl}$ is the classical action we started with \eqref{CS_cl}, $S_{\text{gauge}}$ is the contribution to the BV action due to the usual gauge symmetry and $S_{\text{susy}}$ is the contribution due to the supersymmetry. Denoting the antifield of fields $\phi$ by $\phi^*$ the middle contribution is  
\begin{equation}
   S_{\text{gauge}} = \int d^4x\left( A^*_{\al\dal}D^{\al\dal}c -B^*_{\al\be}[B^{\al\be},c] + \psi^*_{\dal}[c,\psi^{\dal}] + \tilde \psi^*_{\al}[c,\tilde\psi ^{\al}] - \frac{1}{2}c^*[c,c]\right)
\end{equation}
Since the SUSY only closes on-shell, there are second order terms in the antifields in BV action \cite{Baulieu:1990uv}. The twist we will consider is given by $\bep = 0$ and $\vep \neq0$ and constant, and in this case the second order terms in the antifields vanish leaving a simple first-order contribution of the form $\phi^* \delta\phi$:
\begin{equation}
    S_{\text{susy}}= \int d^4x\left(A^*_{\al\dal}\vep^{\al}\psi^{\dal} + \bpsi^*_{\al} \vep_{\be}B^{\al\be} \right)
\end{equation}

After the twist we can define new ghosts numbers which might change the characters of fields and anti-fields. Take the SUSY transformation 
\begin{equation}
\del A_{\al\dal} = \vep_{\al}\psi_{\dal}
\end{equation}
the field $A$ has ghost number zero and $\del$ increases the ghost number by one. That means $\psi$ must have ghost number one after the twist and $\tilde \psi$ must have ghost number $-1$ since the Lagrangian has ghost number zero. That is, the fermionic term of the  classical action in the untwisted theory $\psi^{\dal}D_{\al\dal}\bpsi^{\al}$ is no longer interpreted as a classical term, since it contains terms with ghost numbers different from zero. Following the same logic, the antifield $\psi^*$ must have ghost number zero and is now a physical field. Hence we see that after the twist, the term $\bpsi^*_{\al}\vep_{\be}B^{\al\be}$ must be considered as part of the classical action
which is then 
\begin{equation}
    S_{cl}^{\text{twist}} =\int d^4x\left(B_{\al\be}F^{\al\be} + \frac{g^2}{2}B^2+  \bpsi^*_{\al} \vep_{\be}B^{\al\be}\right).
\end{equation}
To show that this is the holomorphic BF theory, we must rewrite the fields in terms of holomorphic forms. For this, we use the complex structure induced by $\vep_{\al}$ to decompose $B_{\al\be} \in \Omega^{2,0}\oplus \omega\Omega^0\oplus\Omega^{0,2}$, where $\omega$ is the Kahler form 
\begin{equation}
\omega=\epsilon_{\dal\dbe}\vep_{\al}\vep^*_{\be}\,d x^{\dal\al}\wedge dx^{\be\dbe}=d \bz^{\dal}\wedge d z_{\dal}.
\end{equation}
Each component of this decomposition can be obtained from $B_{\al\be}$ by contracting with $\vep^\al$ and its conjugate $\vep^{*\al}$. 
A similar decomposition is performed on $F_{\al\be}$ which also has components in $\Omega^{2,0}\oplus \omega\Omega^0\oplus\Omega^{0,2}$. The fermion $\overline{\psi}^*$ appears in the action as the combination $\overline{\psi}^*_{\al}\vep_{\be} \in \Omega^{2,0}\oplus\omega\Omega^0$, which allows for a field redefinition
\begin{equation}
    \ov{\psi}^*_{\al}\vep_{\be}\rightarrow  \ov{\psi}^*_{\al}\vep_{\be}-\frac{g^2}{2}(B^{1,1}+2B^{2,0}) - F^{1,1} - F^{2,0},
\end{equation}
that eliminates the $B^2$ term and all but the anti-holomorphic part of the connection $F^{0,2}$. The fermion $\overline{\psi}^*$ doesn't couple to $B^{2,0}$ and has no kinetic term, so it can be safely integrated out resulting in the action
\begin{equation}
S_{cl}^{\text{twist}}=\int B^{2,0}\wedge F^{0,2}.
\end{equation}
Note that since the $B^2$ term is eliminated in the supersymmetric twist we have that, perturbatively, the twist of the $\N=1$ sYM is equivalent to the twist of the \textit{self-dual} $\N=1$ sYM. Both can be seen as different gauge fixings of the same twisted BRST complex. 

\subsubsection{$\N\geq2$}\label{subsubsec:N=4}
The holomorphic twist of the vector multiplets for theories with extended supersymmetry follows in a simple way from the minimal twist of $\mathcal{N}=1$, we just need to compute the twist of a chiral multiplet and add the result to the twist of the vector multiplet of $\N=1$ that we found above. That is, since we can describe the vector multiplet of $\N=2$ ($\N=4$) as the vector multiplet of $\N=1$ plus one (three) chiral multiplet(s), we can do the twist separately and sum both results. This will not be true on twistor space, due to extra gauge symmetries present in twistor space, as we will see on section (\ref{sec:TwistOnTwistorSpace}). 

The same procedure from last section can be straightforwardly applied to matter multiplets  \cite{Elliott:2020ecf}. Since the steps are very similar we will review this briefly. The action for the chiral multiplet is
\begin{equation}
    S = \int d^4x\left( D_{\al\dal}\phi D^{\al\dal}\ov{\phi} + \bchi^{\dal}D_{\al\dal}\chi^\al + \text{interactions}\right).
\end{equation}
Here, $\phi$ and $\ov{\phi}$ are the scalars of the chiral multiplet, while $\chi^\al,\bchi^{\dal}$ are the fermions. The interaction terms will drop out of the classical action after the twist so we omit them. 
To perform the twist, we again set $\bep=0$, and consider the following transformations
\begin{equation}
    \del \phi = \vep^\al \chi_\al \quad \del\ov{\phi} = 0 \quad \del \chi_\al = 0 \quad \del \bchi_{\dal} = \vep^\al D_{\al\dal}\ov{\phi}.
\end{equation}
By adding $\del$ to the BRST operator, we find the following grading
\begin{equation}
    \text{gh}[\phi] = 0\quad \text{gh}[\ov{\phi}] = 0 \quad \text{gh}[\chi] = 1 \quad \text{gh}[\ov{\chi}] = -1.
\end{equation}
The classical action after the twist is 
\begin{equation}
    S_{\text{cl,twist}} = \int d^4x\left(\ov{\chi}^{\dal*}\vep^\al D_{\al\dal}\overline{\phi} +D_{\al\dal}\phi D^{\al\dal}\ov{\phi}\right)
\end{equation}
Now, we can perform the holomorphic decomposition:
\begin{equation}
    \ov{\chi}^{\dal*}\vep^\al = \beta  ,\quad \be \in\Omega^{2,1},\quad D_{\al\dal} = D+\ov{D} 
\end{equation}
and eliminate the last term by a field redefinition. The resulting action is
\begin{equation}
    S = \int \be_{2,1}\ov{D}\,\ov{\phi},
\end{equation}
which is a $\be\ga$ system. 

Since the vector multiplet of $\N=4$ can be written as the vector multiplet of $\N=1$ plus three chiral multiplets, the twist of $\N=4$ in four dimensions is a holomorphic BF theory and three $\be\ga$ systems. It was shown in \cite{Costello:2016mgj,Baulieu:2010ch} that this can be written as a holomorphic Chern-Simons on $\mathbb{C}^{2|3}$:
\begin{equation}
    S = \int d^2z d^3\eta\left(\frac{1}{2}\A\bp\A + \frac{1}{3}\A^3 \right),
\end{equation}
where $\A \in \Omega^{0,*}(\mathbb{C}^{2|3})$. By expanding this action, we find the holomorphic BF theory and three $\be\ga$ systems. To see this, expand the connection into definite antiholomorphic form degrees:   
\begin{equation}
    \mathcal{A} = \mathcal{B} + A + \mathcal{B}^*,
\end{equation}
with a scalar $\mathcal{B}$, one-form $A$ and two-form $\mathcal{B}^*$. The action is then
\begin{equation}
    S=\int d^2z \,d^3\eta\left\{ \mathcal{B}\left(\bp A + [A,A]\right) + \mathcal{B^*}[\mathcal{B},\mathcal{B}]\right\}.
\end{equation}
These fields have the following expansions in the fermionic coordinates:
\begin{align*}
    &\mathcal{B} = c + \chi^i\eta_i + \frac{1}{2}\vep^{ijk}\ga_i\eta_j\eta_k + \frac{1}{3}B\eta^3 \quad A = a + \be^i\eta_i + \frac{1}{2}\vep^{ijk}\be^*_i\eta_j\eta_k + \frac{1}{3}a^*\eta^3\\&\mathcal{B}^* = B^* + \ga^{i*}\eta_i + \frac{1}{2}\vep^{ijk}\chi^*_i\eta_j\eta_k + \frac{1}{3}c^*\eta^3.
\end{align*}
Plugging this expansion on the action and performing the fermionic integrals gives the following action:
\begin{equation}
\begin{aligned}
        \int d^2z( & B\bp_a a + \be_i\bp_a\ga^i+\be_i[\chi_j,\be_k]\ep^{ijk}+\beta^{*i}\bp_a\chi_i +B^*[\ga^i,\chi_i]+ \ga^{*}_i[\chi_j,\chi_k]\epsilon^{ijk} \\&+ B[c,B^*] + \beta^{*i}[c,\beta_i] + \ga_i^*[c,\ga_i] + a^*\bp_ac + \chi^{*i}[\chi_i,c] -\frac{1}{2} c^*[c,c]).
\end{aligned}
\end{equation}
The first three terms encode the classical action, which is a BF-term with three $\be\ga$ systems and the auxiliary term $\be_i[\chi_j,\beta_k]$. The terms linear on antifields give the gauge redundancies of this action.
Self-dual $\N=4$ can be written as holomorphic Chern-Simons on $\mathbb{PT}^{3|4}$, so, to reproduce the action above we expect that the twist on twistor space localizes the theory to spacetime and kills one fermionic direction. This is precisely what we will find in the next section.


Theories with extended supersymmetry have more supercharges and a richer structure of possible twists. For example, with at least $\N=2$, the twist with respect to the supercharge $Q = Q_1+Q_2$, gives the \textit{Donaldson-Witten twist} \cite{Witten:1988ze} which is a topological field theory. Another example are the chiral algebras of \cite{Beem:2013sza} for superconformal theories with at least $\N\geq2$ superconformal symmetry, which appear from twisting with the combination of supercharges
\begin{equation}
    \del = Q_1 + \tilde S_2.
\end{equation}
These chiral algebras live on a plane inside the original spacetime, and since they are chiral, depend on the complex structure on this plane. We will see in section \ref{sec:chiralAlgebra} how the chiral algebras can be recovered from an analogous twist in twistor space, which also provides a geometric interpretation for the plane where the chiral algebras live.

\section{The supersymmetric twist on twistor space}\label{sec:TwistOnTwistorSpace}

In this section we perform the supersymmetric twist of holomorphic theories in twistor space using the same tools as in the previous section. Since these theories describe the self-dual sector of Yang-Mills coupled to matter we expect the that their twist is equivalent to the spacetime supersymmetric twist of full Yang-Mills. The twist of pure $\mathcal{N}=1$ in twistor space was performed in \cite{costello2013notessupersymmetricholomorphicfield} in a rather abstract way. Here We will perform the minimal twist for $\mathcal{N}=1$, as well as theories with extended supersymmetry and their matter multiplets in a very explicit way. We will also study the twist that gives rise to the chiral algebras of \cite{Beem:2013sza}, and, for completeness what we call the anti-holomorphic twist, which doesn't seem to correspond to a spacetime twist. Finally, we briefly study the twist of the non-linear graviton describing self-dual $\N=1$ supergravity.

The main difference in twistor space is that holomorphic theories in twistor space have more gauge redundancy than their spacetime counterparts\cite{Boels:2006ir}. This is expected since twistor space is bigger than spacetime, so we need an equally big redundancy in order to describe the same number of degrees of freedom. The fields on twistor space that correspond to spacetime fields can be taken to be represented by Dolbeault cohomology classes, that is, they have gauge redundancies of the form $\del\Phi = \ov{D}\al$, where $\Phi$ represents any component field of the multiplet. The fact that all fields are defined up to $\ov{D}$ exact terms is a reflection that the physical information on twistor space is encoded in cohomology groups. As we shall see, after the twist, the ghosts of some of these gauge symmetries will become physical fields. 
Another consequence of this extra gauge redundancy is that, unlike in spacetime, the supersymmetry transformations we use for the twist do not commute with all gauge transformations. This introduces extra terms to the BV action due to non-trivial structure of the twisted gauge algebra. On top of that the twisted gauge algebra is different between theories with different amounts of supersymmetry and each case has to, at least initially, be treated separately.

Given all of these extra complications, we will first discuss the BV action of the holomorphic $BF$ action on $\mathbb{PT}^{3|1}$ (\ref{N=1_Action}), corresponding to pure self-dual $\mathcal{N}=1$ Yang-Mills. This will serve as a template for the computations in the other theories.

\subsection{The BV action of $\mathcal{N}=1$}\label{sec:BVactionN=1}

The $\N=1$ action (\ref{N=1_Action}) in twistor space
\begin{equation}
    S=\int D^3Z\left( b\ov{D}a + \chi \ov{D}\psi\right).
\end{equation}
has the usual gauge redundancies, analogous to the spacetime $BF$ action 
\begin{equation}\label{usualGauge}
    \del a = \ov{D}c \quad \del B = [c,B] + \ov{D} \xi \quad \del \psi = [c,\psi] \quad \del \chi = [c,\chi],
\end{equation}
where $c$ is the ghost of usual gauge symmetry and $\xi$ is the ghost from the transformation of $B$. But these don't exhaust the redundancies of the action, it is easy to check that the action is also invariant with respect to the transformations:
\begin{equation}\label{extraGauge}
    \del \psi = \ov{D}\Gamma \quad \del\chi = \ov{D}\Lambda \quad \del b = [\Gamma,\chi] - [\psi,\Lambda],
\end{equation}
where $\Lambda$ and $\Gamma$ are ghosts for this extra gauge redundancy. In terms of the superfields (\ref{N=1FIelds}) the classical action is
\begin{equation}
S=\int D^{3|1}Z\,\mathcal{B}\left(\ov{\partial}\mathcal{A}+\mathcal{A}\wedge\mathcal{A}\right)
\end{equation}
The BV action in this case is simple to find, we extend the fields $\mathcal{B}$ and $\mathcal{A}$ to all antiholomorphic form degrees \cite{Baulieu:1995bq}:
\begin{equation}
    \mathcal A \in \Omega^{0,*}(\mathbb{PT}^{3|1},\mathfrak{gl}_N(\mathbb{C}))\quad \mathcal B \in \Omega^{0,*}(\mathbb{PT}^{3|1},\mathfrak{gl}_N(\mathbb{C})).
\end{equation}
Promoting them to $(0,*)$-polyforms, with each component having ghost number obeying $p+g=1$, where $p$ is the form degree and $g$ the ghost number. Explicitly, the expansion in the form degree is
\begin{equation}
    \mathcal{A} = \mathbf{c} + \mathbf{A} + \mathbf{B}^* + \mathbf{\xi}^* \quad \mathcal{B} = \mathbf{\xi} + \mathbf{B} + \mathbf{B^*} + \mathbf{c}^*,
\end{equation}
where $\{\mathbf c, \mathbf\xi\}\in \Omega^{0,0}(\mathbb{PT}^{3|1})$ are the ghosts for the gauge redundancies and $\{\mathbf A, \mathbf B\}\in \Omega^{0,1}(\mathbb{PT}^{3|1})$ the physical fields. The other fields, denoted with $*$ are the antifields, which can be seen as sources for the gauge transformations. All of the components are functions of $Z^A$ and $\eta$. In terms of these the BV-action is
\begin{equation}\label{BVofBF}
    S = \int D^{3|1}Z(\mathbf B\ov{\mathbf D} \mathbf A +\mathbf A^* \ov{\mathbf D}\mathbf c + \mathbf B^*\left(\ov{\mathbf D}\mathbf\xi + [\mathbf c,\mathbf B]\right) - \frac{1}{2}\mathbf c^*[\mathbf c, \mathbf c] + \mathbf \xi^*[\mathbf c,\mathbf \xi])
\end{equation}
Where here, and in the following, we will leave the wedge product of forms implicit. We can go one step further and expand the fields in the fermionic coordinates
\begin{align}
    &\mathbf c = c + \Gamma \eta\quad \mathbf \xi = \Lambda + \xi\eta\quad  \mathbf{A} = a + \psi \eta \quad \mathbf{B} = \chi + b \eta\\& \mathbf A^* = \psi^* + a^*\eta\quad \mathbf B^* = b^* +\chi^*\eta \quad \mathbf c^* = \Gamma^* + c^*\eta \quad \xi^* = \xi^*+\Lambda^*\eta.
\end{align}
where $\Gamma$ and $\Lambda$ are the bosonic ghosts of the extra gauge redundancy (\ref{extraGauge}). The fully expanded BV-action is then
\begin{multline}\label{BVaction}
    S = \int D^3Z( b\ov{D}a + \chi\ov{D}\psi + a^*\ov{D}c + b^*\left([c,b] + [\Gamma,\chi] - [\psi,\Lambda] +\ov{D}\xi\right) +\psi^*\left(\ov{D}\Gamma +[c,\psi]\right)+\\\chi^*\left(\ov{D}\Lambda +[c,\chi]\right)-\frac{1}{2}c^*[c,c] + \Gamma^*[\Gamma,c] + \Lambda^*[\Lambda,c]+\xi^*\left([\xi,c]+[\Gamma,\Lambda]\right)).   
\end{multline}
The first two terms reproduce the classical action, the following terms are linear in the antifields and multiply the gauge transformation of the fields. The terms with  antifields of the ghosts, $c^*,\xi^*,\Gamma^*$ and $\Lambda^*$, contain information about the structure constants of the gauge algebra. 
The BV actions for $\N=2,4$ can be found in a similar way, we write them in Appendix \ref{app:BVaction}.

\subsection{$\N=1$ vector multiplet}\label{sec:N=1vecMultiplet}

The SUSY transformations of \eqref{N=1_Action} are $\del = i\vep^{\al}Q_{\al}+i\bep_{\dal}\overline{Q}^{\dal}$, where $Q$ and $\tilde Q$ are the supersymmetry generators in (\ref{FirstBasis}). Its action on the component fields is:
\begin{align}
    &\del a = \vep^{\al}\lambda_{\al}\psi \quad \del \psi = \bep^{\dal}\p_{\dal}a\\&\del \chi = \vep^{\al}\lambda_{\al}b \quad \del b = \bep^{\dal}\p_{\dal}\chi,
\end{align}
where $\p_{\dal} = \dfrac{\p}{\p\mu^{\dal}}$. We'll first be interested in the $\bep=0$ twist, the other twist will be analyzed later. Lifting it to a gauge redundancy we find that, in contrast to the spacetime twist of $\N=1$, the gauge algebra and the SUSY algebra do not commute. Denote $\del_c,\del_\xi,\del_\Lambda,\del_\Gamma$ the gauge transformation with their corresponding ghost in the subscript. For, $\bep = 0$, we find the following commutation relations
\begin{equation}
    [\del_{\text{susy}},\del_\Gamma]=\del_{c=-\vep^\al\lambda_\al \Gamma}\quad [\del_{\text{susy}}, \del_\xi]=\del_{-\Lambda=\vep^\al\lambda_\al \xi}\quad[\del_{\text{susy}},\del_c]=0 \quad [\del_{\text{susy}},\del_\Lambda]=0.
\end{equation}
This means that The BV action for the twisted complex must have terms that account for this non-trivial gauge algebra. Following the usual BV procedure, we need to add the following terms to the BV action:
\begin{equation}
    -\int D^3Z(c^*\vep^\al\lambda_\al \Gamma+\Lambda^*\vep^\al\lambda_\al \xi).
\end{equation}
The first term accounts for the nontrivial commutator $[\del_{\text{susy}},\del_\Gamma]$, while the second accounts for the non-zero $[\del_{\text{susy}},\del_\xi]$. We also need to add to the BV action the twsiting susy transformations. In total, the terms added to the BV action due to the twist are
\begin{equation}\label{bvSusy}
    S_{\text{susy}} = \int D^3Z(a^*\vep^{\al}\lambda_\al \psi + \chi^*\vep^\al\lambda_\al b -c^*\vep^\al\lambda_\al \Gamma-\Lambda^*\vep^\al\lambda_\al \xi).
\end{equation}
Here, $a^*\in \Omega^{0,2}(\mathbb{PT},\mathcal{O}(0))$, $\chi^*\in \Omega^{0,2}(\mathbb{PT},\mathcal{O}(-1))$,$c^*\in \Omega^{0,3}(\mathbb{PT},\mathcal{O}(-4))$ and $\Lambda^*\in \Omega^{0,3}(\mathbb{PT},\mathcal{O}(-1))$.
From this BV action we identify the new classical action by isolating the fields which have ghost number zero after the twist. Since $a$ and $b$ have ghost number zero, $\psi$ and $\chi$ must have ghost number $1$ and$-1$ respectively. Hence $\chi^*$, which was originally an antifield, now has ghost number zero making it a physical field in the twisted theory.
From the BRST transformation $\del \chi = D\Lambda$ we also have that $\Lambda$ also has ghost number zero, and is also a physical field after the twist. The classical action of the twisted theory is then:
\begin{equation}
    S_{\text{twisted}} = \int D^3Z\Big[b\ov{D}a + \chi^*\left(\vep^\al\lambda_\al b + \ov{D}\Lambda\right)\Big].
\end{equation}
It is worthwhile to mention how the fields transform under each gauge transformation after the twist. They can be read directly from the terms linear in the antifields in the twisted BV action. We see that $\psi$ is the ghost of the gauge transformation $\del_\psi$, that changes the fields by:
\begin{equation}\label{gaugePsi}
    \del_\psi a = \vep^\al\lambda_\al \psi \quad \del_\psi b = -[\psi,\Lambda]\quad \del_\psi\chi^* = -\ov{D}\psi.
\end{equation}
Under $\del_c$, $a$ and $b$ transform in the same way as before, while $\chi^*$ and $\Lambda$ transform as elements of the adjoint of the gauge group. Under $\del_\xi$, the fields transform as
\begin{equation}\label{NewGaugeOfXi}
    \del_\xi b = \ov{D}\xi \quad \del_\xi \Lambda = -\vep^\al\lambda_\al \xi.
\end{equation}
We also note that $\Gamma$ has ghost number two after the twist. This means that $\Gamma$ is now the ghost-for-ghost of $\psi$. Given the gauge transformation (\ref{NewGaugeOfXi}), we can see that $\Lambda$ can be set to zero everywhere except when $\vep^\al\lambda_\al = 0$. This is the first indication that the action is localized to a point on the twistor sphere. To show that the same happens to the other terms in the action, we note that, except at the point $\vep^\al\lambda_\al =0 $, we can perform the field redefinition
\begin{equation}\label{FieldRedefinition}
\chi^*\vep^\al\lambda_\al\rightarrow\chi^*\vep^\al\lambda_\al - \ov{D}a.
\end{equation}
which eliminates the $b\ov{D}a$ term from the action. The leftover term proportional to $\chi^*$ has no kinetic term and can be integrated out, obtaining a trivial classical action $S=0$. This is true almost everywhere except when $\vep^\al\lambda_\al =0$. At this point in the twistor sphere $\mathbb{CP}^1$ the second term vanishes and we cannot remove the holomorphic BF term. This means that the twisted action is \textit{topological} except at the point $\langle\vep\lambda\rangle =0$ where it is a \textit{holomorphic} theory. Some more work needs to be done to show that the holomorphic part is indeed equivalent to the holomorphic BF theory in spacetime.

We can think about the classical action as being localized at a point on the sphere
\begin{equation}\label{N=1Localized}
    S_{\text{twisted}} = \int D^3Z\, \left(b\ov{D}a+\chi^*\ov{D}\Lambda\right)\delta(\langle\lambda\vep\rangle).
\end{equation}
where the delta-function should be taken as projectively identifying both holomorphic and anti-holomorphic components \cite{Adamo:2013cra} and is of weight zero. A choice of point on the twistor sphere $\CP^1$ fixes a, projective constant spinor, which is equivalent to a choice of complex structure on $\mathbb{R}^4$. It is then expected that the action above reduces to a holomorphic BF theory on $\mathbb{C}^2$ where the complex structure dictated by $\vep$. We can show that is the case by a direct reduction to spacetime following \cite{Boels:2006ir,Woodhouse:1985id}. We write the fields in and adapted basis of one-forms (\ref{OneFormsBasis})
\begin{equation}
 a= a_0\ov{e}^0 + a_{\dal}\ov{e}^{\dal},\quad b= b_0\ov{e}^0 + b_{\dal}\ov{e}^{\dal} \quad \chi = \chi_{0\dal}\,\ov{e}^0\wedge\ov{e}^{\dal} + \chi_{\dal\dbe}\,\ov{e}^{\dal}\wedge\ov{e}^{\dbe}
\end{equation}
and pick and auxiliary metric on the twistor sphere to impose the partial gauge fixing
\begin{equation}
    \bp^*a_0|_{{\CP^1}_x} = 0,\quad \bp^* b_0|_{{\CP^1}_x} = 0
\end{equation}
along each $\CP^1$ fiber labeled by their base point $x$. This gauge choice is known as the Woodhouse gauge \cite{Woodhouse:1985id} and we'll make use of it throughout this work. This condition imposes that $a_0$ and $b_0$ must be globally holomorphic functions of weight $0$ and $2$, and so are given by 
\begin{equation}\label{WoodGaugeAnB}
    a_0 = 0 \quad b_0 = B_{\al\be}\lambda^\al\lambda^\be.
\end{equation}
The gauge transformation $\del_\psi$ can be used to eliminate the $\chi^*\ov{D}\Lambda$ term from the action. To eliminate the $\chi_{0\dal}^*$ component consider a gauge transformation with a parameter where $\psi_{0} = 0$, so that
\begin{equation}
    \del\chi^*_{0\dal} = \bp_0\psi_{\dal}.
\end{equation}
Note that $\chi^*_{0\dal}$ is a section of $\mathcal{O}(2)$, while $\psi_{\dal}$ has no holomorphic weight. To analyze how much we can gauge fix the fields, it is useful to expand the fields in weighted spherical harmonics, that is, in polynomials of $u^\al$ and $\hat{u}^\al$, defined by
\begin{equation}
    u_\al=\lambda_\al\quad \hat{u}_\al = \hat{\lambda}_\al/\langle\lambda\hat{\lambda}\rangle.
\end{equation} 
Hence, we can write the fields as
\begin{equation}
    \chi^*_{0\dal} = \sum_{n=0}^{\infty} \chi^*_{0\dal,\al_1..\al_{n+2}\be_1..\be_n}u^{\al_1}..u^{\al_{n+2}}\hat{u}^{\be_1}..\hat{u}^{\be_n},
\end{equation}
\begin{equation}
    \psi_{\dal} = \sum_{n=0}^{\infty} \psi_{\dal,\al_1..\al_{n+2}\be_1..\be_n}u^{\al_1}..u^{\al_{n}}\hat{u}^{\be_1}..\hat{u}^{\be_n}.
\end{equation}

On these variables we can write the anti-holomorphic exterior derivative (\ref{ExteriorDerivativeBasis}) as $\quad \bp_0 = u^\al \frac{\p}{\p\hat u^\al}$. Hence the gauge transformation can be written as
\begin{equation}
    \bp_0\psi_{\dal} = \sum_{n=0}^{\infty} (n+1)\psi_{\dal,\al_1..\al_{n+2}\be_1..\be_n}u^{\al_1}..u^{\al_{n+2}}\hat{u}^{\be_1}..\hat{u}^{\be_n}.
\end{equation}
We find that it is possible to set $\chi_{0\dal}$ to zero, by choosing the gauge transformation  parameters to be
\begin{equation}
    \psi_{\dal,\al_1...\al_{n+2}\be_1..\be_n} = -\frac{1}{n+1}\chi^*_{0\dal,\al_1..\al_{n+2}\be_1..\be_n}.
\end{equation}
With this choice, the leftover contribution from $\chi^*$ to the action is
\begin{equation}
    S_{\text{cl}}\supset \int D^3Z\,\chi^*_{\dal\dbe}\ov{\p}_0\Lambda.
\end{equation}
Since there is no other dependence of $\chi_{\dal\dbe}$ in the action, we can integrate it out keeping in mind that the action is localized to $\lambda=\vep$.  We are thus left with only the BF term: 
\begin{equation}
    S_{\text{twisted}} = \int \frac{D^3Z\wedge D^3\ov{Z}}{\langle\lambda\hat{\lambda}\rangle^4}\Big(\lambda_\al\lambda_\be B^{\al\be}(\bp_{\dal}a_{\dbe}+[a_{\dal},a_{\dbe}]) + b_{\dal}\bp_0a_{\dbe}\Big)\delta(\langle\lambda\vep\rangle).
\end{equation}
The field $b_{\dal}$ is a Lagrange multiplier that can be integrated out to give 
\begin{equation}\label{SpacetimeGaugeField}
    a_{\dbe} = \vep ^\be A_{\be\dbe}
\end{equation}
The volume form can be written as: 
\begin{equation}
    \frac{D^3Z\wedge D^3\ov{Z}}{\langle\lambda\hat{\lambda}\rangle^4} = \frac{\langle\lambda d\lambda\rangle \langle\hat{\lambda}d\hat{\lambda}\rangle}{\langle\lambda\hat{\lambda}\rangle^2} \frac{dx^{\al\dal}dx^{\be\dbe}dx^{\ga\dga}dx^{\del\dot{\del}}}{\langle\lambda\hat{\lambda}\rangle^2}\vep_{\dal\dbe}\vep_{\dga\dot{\del}}\lambda_\al\lambda_\be\hat{\lambda}_\ga\hat{\lambda}_\del
\end{equation}
where we recognize the first term as the measure on $\mathbb{CP}^1$. The integration over the $\mathbb{CP}^1$ fiber is done against the delta function giving:
\begin{equation}
    S_{\text{twisted}} = \int dz_{\dbe}dz^{\dbe} d\ov{z}_{\dal}d\ov{z}^{\dal} \,\vep^{\al}\vep^{\be}B_{\al\be}\vep^\ga\vep^\rho D_{\ga\dga}A^{\dga}_\rho,
\end{equation}
where we have defined
\begin{equation}
    dz^{\dal} =\frac{1}{\langle\vep\hat{\vep}\rangle^{1/2}} \hat{\vep}_\al dx^{\dal\al}\quad d\ov{z}^{\dal} =\frac{1}{\langle\vep\hat{\vep}\rangle^{1/2}} {\vep}_\al dx^{\dal\al}.
\end{equation}
Using the Schouten identity, we recognize this action as
\begin{equation}
    S_{\text{twisted}} = \int_{\mathbb{C}^2} B \ov{D}a,
\end{equation}
with the definitions
\begin{equation}\label{IdentificationsBF}
    B = dz_{\dbe}dz^{\dbe}B^{\al\be}\vep_{\al}\vep_{\be} \quad \ov{D} = \vep^{\ga}d\ov{z}^{\dga}D_{\ga\dga} \quad a = \vep^\al d\ov{z}^{\dal}A_{\al\dal}.
\end{equation}
This is precisely the holomorphic $BF$ theory on $\mathbb{C}^2$ as expected.

\subsection{$\N=2$ vector multiplet}\label{subsec:N=2twist}
We know do the same procedure for $\N=2$. In spacetime, this can been done by performing the twist of the $\mathcal{N}=1$ chiral multiplet and the twist of the $\N=1$ vector multiplet independently. In twistor space this cannot be done a priori due to the extra gauge symmetries. The extra gauge symmetries from $\N=1$ (\ref{extraGauge}), are symmetries of the full action, not of each term individually. To compute the holomorphic twist of $\N=2$ self-dual Yang-Mills we again consider the full BV action in twistor space, see appendix (\ref{BVN=2}), which we repeat here for convenience
\begin{equation}
    S=\int D^3Z\Big( b\ov{D}a + \chi_i\ov{D}\psi^i + \ov{\phi}\,\ov{D}\phi + \frac{1}{2}\ep_{ij}\ov{\phi}\wedge\psi^i\wedge\psi^j\Big).
\end{equation}
For $\N=2$ there are six extra gauge redundancies, one for each physical field.
They are
\begin{equation}\label{N=2Extragauge}
\begin{aligned}
&\del \psi^i = \ov{D}\Gamma^i 
 \quad\del \ov{\phi} = \ov{D}\ov{\sigma} 
\quad \del \chi_i = \ov{D}\Lambda_i - [\psi_i,\ov{\sigma}] + [\Gamma_i,\ov{\phi}] \\
&\del \phi = \ov{D}\sigma + [\Gamma_i,\psi^i]
\quad \del b = \ov{D}\xi + [\Gamma^i,\chi_i] -[\psi^i,\Lambda_i] +[\ov{\phi},\sigma]-[{\phi},\ov{\sigma}],
\end{aligned}
\end{equation}
where $\Gamma^i$ and $\Lambda_i$ are bosonic ghosts while $\sigma$ and $\ov{\sigma}$ are fermionic ghosts. They are all scalars with the same weight as the corresponding field. Note that the gauge transformation of $b$ depends on the ghosts of the $\mathcal{N}=1$ chiral multiplet. That's why, in principle at least, we can't just twist the chiral multiplet and add it to the result in the previous section.

To perform the supersymmetric twist we follow the same procedure as in the previous section. The supersymmetry transformation can be read from the generators (\ref{FirstBasis}) and the component fields of $\N=2$ (\ref{N=2FieldsVector}). For $\del = \vep_i^\al Q_\al^i +\bep^{\dal i}\tilde Q_{\dal i}$ with $\bep=0$, we have
\begin{equation}\label{susyOfN=2}
    \del a = \vep^\al_i\lambda_\al\psi^i \quad \del \psi^i =-\vep^{\al i}\lambda_\al \phi \quad \del \ov{\phi} = -\vep^{\al i}\lambda_\al\chi_i\quad \del \chi_i = \vep_i^\al\lambda_\al b.
\end{equation}
Without loss of generality we set $\vep_2^\al = 0$ since we are interested in the minimal twist, and denote $\vep_1^\al = \vep^\al$. This SUSY transformation is promoted to a gauge redundancy by adding the necessary contributions BV action, including the terms reflecting the non-trivial structure constants of the twisted gauge algebra. The added terms are:
\begin{equation}\label{addedN=2Twist}
    S_{\text{susy}}+\int D^3Z\Big( - c^*\vep^\al\lambda_\al\Gamma^1 + \ov{\sigma}^*\vep^\al\lambda_\al \Lambda_2 + \Gamma^{2*}\vep^\al\lambda_\al\sigma -\Lambda^*_1\vep^\al\lambda_\al\xi\Big),
\end{equation}
where $S_{\text{susy}}$ contain terms of the form $\Phi^*\del\Phi$
\begin{equation}
    S_{\text{susy}}=\int D^3Z\Big( a^*\vep^\al\lambda_\al\psi^1+\chi^{1*}\vep^\al\lambda_\al b -\psi^*_2\vep^\al\lambda_\al\phi - \ov{\phi}^*\vep^\al\lambda_\al\chi_2\Big)
\end{equation}
Just like the $\N=1$ case, these terms introduce new gauge redundancies for the classical fields of the twisted theory. From the susy transformations above (\ref{susyOfN=2}), we can read the following ghost numbers after the twist
\begin{equation}
    \text{gh}[\psi^1] =-\text{gh}[\psi^2] = 1\quad \text{gh}[\phi] = 0 \quad \text{gh}[\ov{\phi}] = 0 \quad \text{gh}[\chi_1] = -\text{gh}[\chi_2] = -1. 
\end{equation}
These assignments mean that $\chi_1^*,\psi_2^*,\Lambda_1$ and $\Gamma^2$ have ghost number zero, and are promoted to physical fields of the twisted complex. The new classical action can be read from the BV action (\ref{BVN=2}) together with (\ref{addedN=2Twist}):
\begin{equation}
    S_{\text{cl}} = \int D^3Z\Big(b\ov{D}a + \ov{\phi}\,\ov{D}\phi + \chi_1^*\left(\vep^\al\lambda_\al b +\ov{D}\Lambda_1 -[\Gamma^2,\ov{\phi}]\right) +\psi_2^*\left(\vep^\al\lambda_\al \phi + \ov{D}\Gamma^2\right)\Big).
\end{equation}
Recall that $\chi^*_1\in\Omega^{0,2}(\mathbb{PT},\mathcal{O}(-1))$,$\Lambda\in\Omega^{0,0}(\PT,\O(-3))$ and $\Gamma\in\Omega^{0,0}(\PT,\O(-1))$. From the terms added to the BV action (\ref{addedN=2Twist}), we see that under a gauge transformation $\del_\xi$, the field $\Lambda_1$ has the gauge freedom $\del_\xi\Lambda = -\vep^\al\lambda_\al\xi$ and $\Gamma^2$ has the gauge freedom $\del_\sigma\Gamma =\vep^\al\lambda_\al \sigma$. Hence we can set these fields to zero everywhere except when $\vep^\al\lambda_\al=0$. Then, the field redefinitions
\begin{equation}\label{fieldRedefinitionsTwo}
    \chi^*_1\vep^\al\lambda_\al b \rightarrow \chi^*_1\vep^\al\lambda_\al b -b\ov{Da}\quad \psi_2^*\vep^\al\lambda_\al \phi_2 \rightarrow \psi_2^*\vep^\al\lambda_\al \phi - \ov{\phi}\,\ov{D}\phi.
\end{equation}
can be used to eliminate the other terms in the action. As before, this is valid everywhere except at the point $\vep_\al=\lambda_\al$, where the action becomes
\begin{equation}
    S=\int D^3Z\Big[b \ov{D}a + \ov{\phi}\,\ov{D}\phi + \chi_1^*\left(\ov{D}\Lambda^1 - [\Gamma^2,\ov{\phi}]\right)+\psi_2^*\ov{D}\Gamma^2    \Big]\del\left(\langle\lambda\vep\rangle\right).
\end{equation}
We again use the Woodhouse gauge for a partial gauge fixing along the fibers. The constraints on the fields $a$ and $B$ are the same from the previous section (\ref{WoodGaugeAnB}). For the scalar fields, Woodhouse gauge implies 
\begin{equation}\label{ScalarsN=2Comonents}
    \phi = \phi_0(x)\ov{e}^0 + \phi_{\dal}(x,\lambda,\hat{\lambda})\ov{e}^{\dal} \quad \ov{\phi} = \ov{\phi}_0(x)\ov{e}^0 + \ov{\phi}_{\dal}(x,\lambda,\hat{\lambda})\ov{e}^{\dal}. 
\end{equation}
Plugging (\ref{ScalarsN=2Comonents}) into the kinetic term and integrating out $\phi_{\dal}$ we obtain
\begin{equation}
    \bp_0 \ov{\phi}_{\dal} = \ov{D}_{\dal}\ov{\phi}_0,
\end{equation}
where $\ov{D}_{\dal} = \lambda^\al D_{\al\dal}$ and $D_{\al\dal}$ is the covariant derivative with respect to $A_{\al\dal}$ of (\ref{SpacetimeGaugeField}). The solution for this equation is 
\begin{equation}\label{WoodGaugeScalar}
    \ov{\phi}_{\dal} = \frac{\hat{\lambda}^{\al}}{\langle\lambda\hat{\lambda}\rangle} D_{\al\dal}\ov{\phi}_0.
\end{equation}
In section \ref{sec:N=1vecMultiplet}, we showed that we can use the gauge freedom $\del_{\psi_1}\chi^*_1=\ov{D}\psi_1$ to eliminate the $\chi^*_{0\dal}$ component. This is true because $\chi^*_{0\dal}\in \mathcal{O}(2)$. We can do a similar analysis for the gauge freedom $\del_{\chi}\psi^*_2 = \ov{D}\chi_2$, in particular for the component $\psi_{2,0\dal}^*$, where we have $\del\psi^*_{2,0\dal} =\bp_0\chi_{\dal}+\ov{D}_{\dal}\chi_0$. Writing the expansion in spherical harmonics for $\chi_{\dal}\in\mathcal{O}(-2)$
\begin{equation}
    \chi_{\dal} =\sum_{n=0}^\infty \chi_{\dal,\al_1...\al_n\be_1...\be_{n+2}}u^{\al_1}...u^{\al_n}\hat{u}^{\be_1}...\hat{u}^{\be_n},
\end{equation}
we can see that 
\begin{equation}
    \bp_0\chi_{\dal} =\sum_{n=1}^\infty \chi_{\dal,\al_1...\al_n\be_1...\be_{n}}u^{\al_1}...u^{\al_n}\hat{u}^{\be_1}...\hat{u}^{\be_n}.
\end{equation}
Note that the derivative does not have the $n=0$ term, which is present in the expansion of a $\mathcal{O}(0)$ field. Since $\psi_{2,0\dal}^* \in\mathcal{O}(0)$, we cannot eliminate the mode which is constant along the fibers using the $\bp_0 \chi_{\dal}$ freedom. Hence, we can write $\psi^*_{2,0\dal} = \be_{\dal}(x)$, which is independent of the $\mathbb{CP}^1$ coordinates. Note that we didn't use all the gauge freedom, since we still have the $\bp_{\dal}\chi_0$ term on the gauge transformation. After the reduction to spacetime this becomes the usual gauge freedom of a $\be\ga$ system.

We also have a term on the action of the form $\del(\lambda_\al-\vep_\al)\psi^*_{2,\dal\dbe}\bp_0\Gamma$. After integrating out $\psi_{1,\dal\dbe}^*$ we obtain the constraint
\begin{equation}
    \bp_0\Gamma^2\Big|_{\lambda_\al=\vep_\al} = 0, \Gamma^2\in \O(-1).
\end{equation}
If this were true globally, we would not have a solution for $\bp_0\Gamma^2 = 0$. Since we only need this to be true at $\lambda_\al=\vep_\al$ we have the solution
\begin{equation}
    \Gamma^2 = \frac{\langle\vep\hat{\lambda}\rangle}{\langle\lambda\hat{\lambda}\rangle}\ga(x),
\end{equation}
Now, we need to integrate out the $\chi^{1*}_{\dal\dbe}$ component. This give us the constraint:
\begin{equation}\label{consistency1}
    \bp_0\Lambda\Big|_{\lambda_\al=\vep_\al} = [\Gamma^2,\ov{\phi}]\Big|_{\lambda_\al=\vep_\al} = \frac{\langle\vep\hat \lambda\rangle}{\langle\lambda\hat{\lambda}\rangle}[\gamma,\ov{\phi}_0]\Big|_{\lambda_\al=\vep_\al}.
\end{equation}
Since $\Lambda\in\O(-3)$ we have the following expansion in spherical harmonics  
\begin{equation}
    \Lambda = \sum_{n=0}^\infty \Lambda_{\al_1..\al_n\be_1..\be_{n+3}}u^{\al_1}..u^{\al_n}\hat{u}^{\be_1}..\hat{u}^{\be_{n+3}}.
\end{equation}
Hence
\begin{equation}\label{bp0Lambda}
    \bp_0\Lambda = \sum_{n=1}^\infty (n+2)\Lambda_{\al_1..\al_n\be_1..\be_{n+1}}u^{\al_1}..u^{\al_n}\hat{u}^{\be_1}..\hat{u}^{\be_{n+1}}.
\end{equation}
Since we don't have the mode with $n=0$ in (\ref{bp0Lambda}) and $\Gamma^\al,\ov{\phi}\in \O(0)$, the only solution is that $\Lambda$ and $\ov{\phi}_0$ are constant and can be integrated out. From (\ref{WoodGaugeScalar}) this also sets the kinetic term $\ov{\phi}\,\ov{D}\phi$ to zero. We are left with the following action
\begin{equation}
    S= \int D^3Z\left[b\ov{D}a + \frac{\langle\vep\hat{\lambda}\rangle}{\langle\lambda\hat\lambda\rangle}\be^{\dal}\ov{D}_{\dal}\gamma\right]\delta(\langle\lambda\vep\rangle).   
\end{equation}
Following the same procedure of section \ref{sec:N=1vecMultiplet}, we use the delta function to write the volume form as the volume form of $\mathbb{C}^2$ with the complex structure determined by $\vep_\al$. The action is then written as
\begin{equation}
    S = \int_\mathbb{C^2}b\ov{D}a + \be\ov{D}\ga,
\end{equation}
where we have defined 
\begin{equation}
    \be = \be_{\dal} d\ov{z}^{\dal}dz^{\dbe}dz_{\dbe}  \quad \ov{D} = \vep^\ga d\ov{z}^{\dga}D_{\ga\dga}.
\end{equation}
This is the action for the holomorphic $BF$ theory coupled to a $\be\ga$ system, reproducing the twist on spacetime.


\subsection{$\N=4$ vector multiplet}\label{N=4Holotwist}
The minimal twist of the $\N=4$ vector multiplet on twistor space is very much analogous to the $\mathcal{N}=2$ case. Since there is no additional subtlety we will describe it here only briefly, leaving the detailed computations to Appendix \ref{app:ReductionToSpacetime}. 
The action of self-dual $\N=4$ on twistor space is:
\begin{equation}\label{n=4action}
    S = \int_{\mathbb{PT}}D^3Z\Big( b\ov{D}a + \psi^i\ov{D }\chi_i +\frac{1}{4} \epsilon_{ijkl}\phi^{ij}\ov{D}\phi^{kl} +  \frac{1}{2}  \ep_{ijkl}  \psi^i\psi^j\phi^{kl}\Big).
\end{equation}
To perform the twist, we will write the fields using a $SU(3)\times U(1) \subset SU(4)$ decomposition of the $R$-symmetry group. We pick an $SU(4)$ vector $v_i$ which is left invariant by our choice of $SU(3)\subset SU(4)$, and define the $SU(3)$-invariant tensor
\begin{equation}
    \vep^{IJK} = v_i \vep^{iIJK} \quad \vep_{IJK} = \del^{lm}v_l\vep_{mIJK}
\end{equation}
Where we used $I$ to label the $SU(3)$ directions of R-symmetry components. When doing the twist, we will have $I=2,3,4$. We decompose the fields as:
\begin{equation}
    \psi^j = (v_i\psi^i,\psi^I) \quad \chi_j = (\chi_i v_j\del^{ij},\chi_I) \quad \phi^{ij}=(\Phi^I,\Phi_I) \quad \Phi^I = v_j\phi^{Ij} \quad \Phi_I = \vep_{IJK}\phi^{JK}
\end{equation}

To perform the twist we set $\bep_i = \vep_I = 0$ and $\vep^\al_1 = \vep^\al$, and so only the $v_1$ component of $v_i=1$ is non-zero. The susy relevant transformations are then: 
\begin{equation}
    \del a = \vep^\al\lambda_\al \psi^1 \quad \del \psi^I = \vep^\al\lambda_\al\Phi^{I} \quad \del \Phi_{I}=  \vep^\al\lambda_\al \chi_I \quad \del \chi_1 = \vep^\al\lambda_\al b.
\end{equation}
To twist we add the terms  
\begin{equation}
\begin{aligned}
    \int D^3Z\Big( &a^*\vep^\al\lambda_\al\psi^1 +\psi_I^*\vep^\al\lambda_\al \phi^{1I} +\phi^{1I*}\vep^\al\lambda_\al\chi_I + \chi^{1*}\vep^\al\lambda_\al b\\& -c^*\vep^\al\lambda_\al\Gamma^1-\sigma^{1I*}\vep^\al\lambda_\al \Lambda_I -\Lambda^{1*}\vep^\al\lambda_\al\xi - \Gamma^*_I\vep^\al\lambda_\al \sigma^{1I}\Big)
\end{aligned}
\end{equation}
to the original BV action. The new relevant ghost numbers are 
\begin{equation}
    \text{gh}[\psi^I]=-1 \quad \text{gh}[\psi^1]=1 \quad \text{gh}[\Phi^{I}]=0 \quad \text{gh}[\Phi_{I}]= 0  \quad \text{gh}[\chi_I] = 1 \quad \text{gh}[\chi_1]=-1, 
\end{equation}
with ghost numbers of the other fields fixed by these. By inspection of the BV action (\ref{BVN=4}) we see that the classical action after the twist is 
\begin{equation}\label{TwistedN=4}
    S_{\text{cl,}\N=4} = \int D^3Z\Big(b\ov{D}a + \Phi^{I}\ov{D}\Phi_{I}+ \chi^{1*}\left(\vep^\al \lambda_\al b +\ov{D}\Lambda_1 - [\Gamma^I,\phi_{I}]\right) +\psi^*_I\left(\ov{D}\Gamma^I + \vep^\al\lambda_\al \Phi^{I}\right)\Big).
\end{equation}
which again can be shown to be a $Q$-exact term away from $\vep_\al\lambda^\al=0$. The full reduction to spacetime can be found on Appendix \ref{app:ReductionToSpacetime}, and is completely analogous to the $\N=2$ case. At the end, we find a holomorphic $BF$ theory  coupled to three $\be\ga$ systems depending on the complex structure defined by the spinor $\vep$, matching the twist on spacetime. 

\subsection{The chiral algebra twist}\label{sec:chiralAlgebra}

Now that we have shown that the minimal twist in twistor space reduces to the twisted theory on spacetime, any further compatible supersymmetric twists on twistor space on top of the original one will coincide with the analogous twist on spacetime. But it might still be interesting to understand their structure from the perspective of twistor space. This is the case for the Chiral algebra twist which we examine in this subsection. It is also necessary to understand them in twistor space in order to find their duals in terms of the chiral holography framework.

Here we will assume that the theories we'll be studying have superconformal symmetry, meaning they have extra supersymmetries denoted $S$ apart from the original $Q$ ones. The twist will study is the analogous to the twist in \cite{Beem:2013sza} which the authors used to find an equivalence between certain classes of BPS operators in $D=4$ $\N\geq 2$ superconformal theories and $D=2$ chiral algebras. There, they were interested in the cohomology of an operator of the form $Q+\tilde S$ which is a combination of a regular supersymmetry charge and a superconformal supercharge. They found that the cohomoplogy of this operator localizes to a plane inside $\mathbb{R}^4$ with an induced complex structure, where operators in the cohomology are are holomorphic in this induced complex structure.

We saw above how a similar localization, and induced complex structure, occurs in the minimal twist of SUSY theories in twistor space, and we expect something similar to happen again in this case. By doing the twists on twistor space we expect that after the minimal twist, there are choices of $\mathbb{C}\in\mathbb{C}^2$ which are parametrized by the choice of the second supercharge $S$. This plane will be seen to have a natural geometric interpretation on twistor space.

\subsubsection{$\N=2$}

We will start by finding the chiral algebra corresponding to the twist of the vector multiplet of $\N=2$. As we mentioned on section \ref{subsubsec:N=4}, to find the chiral algebra we need to perform the twist with respect to the superconformal supercharge which is conjugate to the one we did the first twist. The structure of the superconformal twist is very similar to the one of the holomorphic twist. The relevant  superconformal transformations for the twist are
\begin{equation}
    \del_S = \ep_i^{\dal}\tilde S_{\dal}^i + \ep^{\al i}S_{\al i},
\end{equation}
with bosonic spinor parameters $\ep^\al$ and $\epb^{\dal}$.
On the previous sections we twisted the theory with the supercharge $Q_\al^1$ so the relevant superconformal supercharge is $\tilde S_{\dal}^2$. Hence, we set $\ep^{\al i} =0$ and $\epb^{\dal}_1 = 0$, denoting $\epb^{\dal}_2 = \epb^{\dal}$. Recall the form of this generator in twistor space:
\begin{equation}
    \tilde S ^i _{\dal} = \mu_{\dal}\frac{\p}{\p\eta_i}.
\end{equation}
The charge $\tilde{S}$ has the same form as $Q^i_\al$under the exchange $\mu_{\dal }\leftrightarrow\lambda_\al$, which is behind the further localization of the twisted theory. From the form of the generator it generates transformations similar to (\ref{susyOfN=2}), but with the exchange of $\lambda_{\al}$ to $\mu_{\dal}$ and $\vep_1$ to $\epb_2$:
\begin{equation}\label{superconformaTransformation}
    \del a = \epb^{\dal}\mu_{\dal}\psi^2\quad \del \psi^1 =- \epb^{\dal}\mu_{\dal}\phi \quad \del \ov{\phi} =-\epb^{\dal}\mu_{\dal} \chi_1 \quad \del\chi_2 = \epb^{\dal}\mu_{\dal} b.
\end{equation}
Putting together the transformations (\ref{susyOfN=2}) and (\ref{superconformaTransformation}), we find that the twisted ghost number of each field is given by 
\begin{equation}\label{ghostNUmberChiralTwistN=2}
    \text{gh}[a]=\text{gh}[b]=0 \quad \text{gh}[\psi^i]= 1 \quad \text{gh}[\phi] = -\text{gh}[\ov{\phi}] =2 \quad \text{gh}[\chi_i] = -1.
\end{equation}
The terms added to the BV action are the ones from the holomorphic twist (\ref{addedN=2Twist}) plus the analogous terms exchanging $\vep_{1\al}\lambda^\al$ with $\epb_{2\dal}\mu^{\dal}$:
\begin{equation}\label{AddedTermsChiralTwistN=2}
    \int D^3Z\Big( - c^*\epb_{\dal}\mu_{\dal}\Gamma^2 + \ov{\sigma}^*\epb_{\dal}\mu^{\dal} \Lambda_1 + \Gamma^{1*}\epb_{\dal}\mu^{\dal}\sigma -\Lambda^*_2\epb_{\dal}\mu^{\dal}\xi\Big) +S_{\textbf{susy}},
\end{equation}
where again $S_{\text{susy}}$ refers to terms of the form $\Phi^*\del\Phi$ where $\del$ is given by (\ref{superconformaTransformation}).
Hence, $\chi^*_i$ is now a physical field and, consequently, so is $\Lambda_i$. With this new twist the field $\ov{\sigma}^*$ has ghost number zero and also becomes a physical field. Using (\ref{AddedTermsChiralTwistN=2}), we can see that the classical action is: 
\begin{equation}\label{N=2ChiralAction}
    S_{\text{cl}} = \int D^3Z\Big( b\ov{D}a +\chi^*_1\left(b\vep^\al\lambda_\al+ \ov{D}\Lambda_1\right) +\chi^*_2\left(b\epb_{\dal}\mu^{\dal} + \ov{D}\Lambda_2\right) + \ov{\sigma}^* \left(\vep^\al\lambda_\al \Lambda_2 - \epb^{\dal}\mu_{\dal}\Lambda_1\right)\Big).
\end{equation}
We can see that$\sigma^*$ plays the role of a Lagrange multiplier that fixes 
\begin{equation}\label{Ghostconstraint}
    \vep^\al\lambda_\al \Lambda_2 = \epb^{\dal}\mu_{\dal}\Lambda_1.
\end{equation}
Now, we can use the gauge freedom parametrized by $\xi$
\begin{equation}\label{GaugeXiCiralTwistN=2}
    \del_\xi b = \ov{D}\xi \quad \del_\xi\Lambda_1 = \vep^\al\lambda_\al \xi \quad \del_\xi \Lambda_2 = \epb^{\dal}\mu_{\dal}\xi,
\end{equation}
to set $\Lambda_1$ to zero. Given the relation (\ref{Ghostconstraint}), we see that this gauge transformation also eliminates the $\Lambda_2$ term. The only point where this gauge transformation cannot be done is when $\vep^\al\lambda_\al=0$ and $\epb_{\dal}\mu^{\dal} = 0 $. After eliminating these fields a similar field redefinition to (\ref{FieldRedefinition}) can de done to turn the action into a $Q$-exact term, so the physical action is localized to $\vep^\al\lambda_\al = 0$ and $\epb_{\dal}\mu^{\dal} = 0$. The spacetime interpretation of this point can be seen starting from the incidence relation evaluated on the first constraint
\begin{equation}
\mu^{\dal}=x^{\dal\al}\lambda_\al\rightarrow \mu^{\dal}=x^{\dal\al}\vep_\al=z^{\dal}
\end{equation}
where $z^{\dal}$ is the induced complex structure on $\mathbb{R}^4$. On top of it, the second constraint
\begin{equation}
\epb_{\dal}\mu^{\dal}=\epb_{\dal}z^{\dal}=0
\end{equation}
picks out a linearly embedded complex plane in $\mathbb{C}^2$ 
where the chiral algebras of \cite{Beem:2013sza} live.
The non-trivial action can be written schematically as
\begin{equation}
    S = \int D^3Z \left(b\ov{D}a + \chi^*_1\ov{D}\Lambda_1 + \chi^*_2\ov{D}\Lambda_2\right)\,\del(\langle\vep\lambda\rangle)\del([\mu\epb]).
\end{equation}
The reduction to spacetime follows in a similar fashion to the one in \ref{sec:N=1vecMultiplet}, the terms with $\Lambda_i$ can eliminated using the extra gauge symmetry. The leftover action is 
\begin{equation}
    S = \int_{\mathbb{C}^2}B\ov{D}a\,\,\del(\ov{\vep}_{\dal}z^{\dal})
\end{equation}
Because the leftover directions are topologically trivial the reduction to the chiral algebra plane is straightforward. 
In this plane, we are left with a holomorphic BF and the gauge fields can be gauge fixed to zero, producing a $b-c$ ghost system. 
Without the ghost zero mode, which is automatically decoupled from the action, this is reproduces the chiral algebra corresponding to the $\N=2$ vector multiplet found on \cite{Beem:2013sza}.

\subsubsection{$\N=4$}

The same twist for the $\N=4$ theory is very similar to the $\N=2$ and the holomorphic twist detailed in (\ref{TwistedN=4}), so we will be brief.
We use R-symmetry indices $i,a$ and $b$ taking values $i=\{1,2,3,4\}$; $a=\{1,2\}$ and $b=\{3,4\}$. The twisting supersymmetry $\del' =\del_Q +\del_S$ is the same as before, acting as
\begin{equation}
\begin{aligned}\label{susyChiralN=4}
    &\del' a = \vep^\al\lambda_\al \psi^1 +\epb^{\dal}\mu_{\dal}\psi^2 \quad \del'\psi^i= \vep^\al\lambda_\al \phi^{1i}+\epb^{\dal}\mu_{\dal}\phi^{2i} \\& \del' \phi_{ij} =\vep^\al\lambda_\al \del_i^1\chi_j + \epb^{\dal}\mu_{\dal}\del_i^2\chi_{j} \quad \del'\chi_{i} =\vep^\al\lambda_\al \del_i^1 b+ \epb^{\dal}\mu_{\dal}\del_i^2b.
\end{aligned}
\end{equation}
We again add $\del'$ to the BRST operator which induces a change of ghost numbers. Inspecting (\ref{susyChiralN=4}) one can see that the only consistent assignment is
\begin{equation}
\begin{aligned}
    &\text{gh}[\psi^{a}]=1 \quad \text{gh}[\psi^b]= -1 \quad \text{gh}[\phi^{12}]=2\quad\text{gh}[\phi^{ab}]=0 \\&\text{gh}[\phi_{12}]=-2\quad\text{gh}[\phi_{ab}]=0\quad  \text{gh}[\chi_a] = -1 \quad \text{gh}[\chi_b] = 1.
\end{aligned}    
\end{equation}
From the BV action (\ref{BVN=4}), we can read the classical action after the twist as
\begin{multline}\label{N=4ChiralAction}
    S_{\text{chiral }\N=4}=\int D^3Z\Big( b\ov{D}a+ \frac{1}{2}\ep_{ab}\ep_{aa'}\phi^{bb'}\ov{D}\phi^{a'b'} +\chi^{*1}\left(\vep^\al\lambda_\al b +\ov{D}\Lambda_1 +[\Gamma^b,\phi_{1b}]\right)\\+\chi^{*2}\left(\epb^{\dal}\mu_{\dal} b +\ov{D}\Lambda_2+[\Gamma^b,\phi_{2b}]\right) + \sigma^{12*} \left(\vep^\al\lambda_\al \Lambda_2 - \epb^{\dal}\mu_{\dal}\Lambda_1+ \frac{\ep_{bb'}}{2}[\Gamma^b,\Gamma^{b'}]\right)\\ +\psi^*_b\left(\ov{D}\Gamma^b +\vep^\al\lambda_\al \phi^{1b}+\epb^\al\mu_{\dal} \phi^{2b}\right) \Big)
\end{multline}
Where we recognize the action (\ref{N=2ChiralAction}), within (\ref{N=4ChiralAction}), with some extra terms, including the commutators of $\Gamma^b$. The reduction to spacetime follow the same steps as we did in the previous sections and we show the details on Appendix \ref{subapp:ReducChiralN=4}. As expected, the theory is localized to $\langle\vep\lambda\rangle=0,\,[\epb\mu]=0$. In the same way as in the holomorphic twist, when we integrate over the fibers, we obtain 
\begin{equation}\label{ChiralAlgebraN=4Action}
    S = \int_{\mathbb{C}^2}\left(b\ov{D}a + \be^b\ov{D}\ga_b\right)\del\left(\epb_{\dal}z^{\dal}\right),
\end{equation}
where now there are only two symplectic bosons due to the extra twisting supersymmetry. The reduction to the plane is immediate, producing  $b-c$ ghost system coupled to two symplectic bosons, reproducing the results of \cite{Beem:2013sza}.

\subsection{Other twists}

Since we showed that the minimal twist localizes the theory to spacetime, any further supersymmetric twist should give the same  result as the twists on spacetime \cite{Elliott:2020ecf}. A possible interesting twist is with respect to the other supercharge  $\tilde Q$. For this one we don't have an argument that it should coincide with the spacetime twist of the full supersymmetric Yang-Mills. This discrepancy with the twist by $Q$ comes because of the chiral form of the action we started with and can be traced back to the form of that supersymmetry takes in the Chalmers-Siegel action \eqref{CS_cl}. For completeness, we perform the twist by $\tilde Q$ below. Later we also perform the holomorphic twist of the non-linear graviton with $\N=1$ which we expect, but do not show, to coincide with the conjectured twist of pure $\N=1$ sugra \cite{Williams:2020fwd}. 

\subsubsection{Anti-holomorphic twist}

To perform the twist with respect to the supercharge $\tilde Q$ to the $\N=1$ vector multiplet we set $\vep_\al=0$, finding the following susy transformations 
\begin{equation}\label{susyAnti-holo}
    \del \psi =\epb^{\dal}\p_{\dal}a =\tilde \p a\quad \del b = \epb^{\dal}\p_{\dal}\chi=\tilde \p\chi.
\end{equation}
Where we defined the notation $\epb^{\dal}\p_{\dal} = \tilde\p$. As usual, to do the twist, we add the supersymmetry operator to the BRST operator and verify the change in the cohomological grading of the fields. Taking $a,b$ to have ghost number zero after the twist, and given the supersymmetry transformations above, we can find that $\psi$ has ghost number $-1$ after the twist, while $\chi$ has ghost number one. In the same way as in the holomorphic twist, the transformations (\ref{susyAnti-holo}) do not commute with the gauge transformations (\ref{usualGauge}) and (\ref{extraGauge}). This means that we also need to add to the action terms proportional to $[\del_{\textbf{susy}},\del_c], [\del_{\textbf{susy}},\del_\Gamma]$ and $[\del_{\textbf{susy}},\del_\Lambda]$, where $\del_c,\del_\Gamma$ and $\del_\Lambda$ are the gauge transformations of the respective ghost on the subscript. Proceeding in this way, the terms added to the action are: 
\begin{equation}
    \int D^3Z\Big(\psi^* \tilde \p a + b^*\tilde \p \chi + \Gamma^*\tilde \p c + \xi^* \tilde \p \Lambda\Big).
\end{equation}
Looking at the BV action (\ref{BVaction}), $\psi^*$ and $\Gamma$ have ghost degree zero. This implies that the classical action is 
\begin{equation}
    S_{\text{cl}}=\int D^3Z\Big(b\ov{D} a  + \psi^*\left(\tilde \p a + \ov{D \Gamma}\right)\Big).
\end{equation}
Note that, under a gauge transformation, $\Gamma$ changes as
\begin{equation}
    \del_c\Gamma = [c,\Gamma] + \tilde\p c.
\end{equation}
After the twist, $\chi$ has ghost number one; hence it is the ghost of the gauge transformation 
\begin{equation}
    \del_{\chi}b = [\chi,\Gamma]+\tilde \p \chi \quad \del_\chi \psi^* = \ov{D}\chi.
\end{equation}
$\Lambda$ has ghost number two. In the language of \cite{Gomis:1994he}, this reflects the fact that this theory is reducible. Actually, the term $b^* [\psi,\Lambda]$ is quadratic in the antifields (recall that after the twist $\psi$ is the antifield of the classical field $\psi^*$). This is present only in \textit{on-shell first-stage reducible} theories. That is, on-shell, the system has ghost for ghosts. This is reflected by the fact that $\del_\Lambda S_{\text{cl}}$ is zero only on shell. This classical action doesn't localize to spacetime, so it can't correspond to some twist of the full pure $\N=1$ YM.

\subsubsection{Twisting the non-linear graviton}

The action for self-dual Einstein gravity can be written in twistor space as 
\begin{equation}
    S = \int D^3Z\, \tilde h\Big(\bp h +\frac{1}{2}\{h,h\}\Big). 
\end{equation}
Where $\tilde h \in \Omega^{0,1}(\mathbb{PT},\mathcal{O}(-6))$ and $ h \in \Omega^{0,1}(\mathbb{PT},\mathcal{O}(2))$. When reduced to spacetime, this becomes the action for self-dual GR \cite{Mason:2007ct,Wolf:2007tx,Sharma:2021pkl}. The $\N=1$ supersymmetric version can be obtained from a truncation of the $\N=8$ supergravity action of \cite{Mason:2007ct}. This action is composed by a single superfield:
\begin{equation}
    H = h + h^I_1\eta_I + ... +h_{7I}\eta^{7I} +\eta^8\tilde h,
\end{equation}
where $I$ denotes the indices in the fermionic directions and the number as a subscript of $h$ denotes the power of $\eta$ in the expansion. It is easy to see that $h_1^I \in \Omega^{0,1}(\PT,\O(1))$ and $h_{7I}\in \Omega^{0,1}(\PT,\O(-5))$. 
The action functional is
\begin{equation}
    S = \int D^{3|8}Z \,H\Big(\bp H +\frac{1}{2}\{H,H\}\Big).
\end{equation}
Analogously as what was done for Yang-Mills, to find the $\mathcal{N}=1$ action we impose invariance under an $SO(7,\mathbb{C})$ subgroup of the $R-$symmetry group. That is, we consider fields with fermionic components that are either $\eta = \eta_8$ or  $\eta_1\eta_2...\eta_7$.
This gives us two super fields 
\begin{equation}
    \mathbf{h} = h + \eta h_1 \quad \tilde{\textbf{h}} = h_7+\eta\tilde h.
\end{equation}
with $\N=1$ action  
\begin{equation}
    S=\int D^3Z\, \Big(\tilde h\Big(\bp h +\frac{1}{2}\{h,h\}\Big) + h_7\Big(\bp h_1 + \{h,h_1\}\Big)\Big).
\end{equation}
The gauge redundancies are 
\begin{equation}
    \del h =\bp f + \{h,f\} \quad \del \tilde h = \bp \phi + \{h,\phi\} + \{\tilde h,f\} 
\end{equation}
together with
\begin{equation}
    \del h_1 = \bp \Gamma + \{h,\Gamma\}\quad\del h_7 = \bp \Lambda + \{h,\Lambda\}\quad \del \tilde h = \{\Gamma,h_7\}-\{\Lambda,h_1\}
\end{equation}
The BV action of this theory is analogous to (\ref{BVaction}), with the replacement of the commutator by the Poisson bracket. We take as the twisting supersymmetry the transformations
\begin{equation}
    \del h = \vep^\al\lambda_\al h_1 \quad \del h_7 = \vep^\al\lambda_\al \tilde h,
\end{equation}
for which we add the terms 
\begin{equation}
    \int D^3Z\Big(h^*\vep^\al\lambda_\al h_1+h_7^*\vep^\al\lambda_\al \tilde h - f^*\vep^\al\lambda_\al \Gamma -\Lambda^*\vep^\al\lambda_\al\phi\Big).
\end{equation}
to the BV action. Note that $h_7^*\in\Omega^{0,2}(\PT,\O(1))$ and $\Lambda\in\Omega^{0,0}(\PT,\O(-5))$. After the twist $h_7^*$ and $\Lambda$ have ghost number zero, and we identify the new classical action as
\begin{equation}
    S= \int D^3Z\,\Big( \tilde h\Big(\bp h +\frac{1}{2}\{h,h\} \Big)+ h_7^*\left(\vep^\al\lambda_\al \tilde h + \bp \Lambda + \{h,\Lambda\}\right)\Big).
\end{equation}
Using the gauge transformation
\begin{equation}\del_\phi \Lambda = -\vep^\al\lambda_\al\phi
\end{equation}
we can eliminate the $\Lambda$ field everywhere, except at $\langle\vep\lambda\rangle=0$. Doing a field redefinition:
\begin{equation}
    h^*_7\vep^\al\lambda_\al\tilde h\rightarrow h^*_7\vep^\al\lambda_\al\tilde h - \tilde h \left(\bp h + \frac{1}{2}\{h,h\} \right),
\end{equation}
we can fully localize the action to a point in the twistor sphere 
\begin{equation}
    S= \int D^3Z\, \tilde h\Big(\bp h +\frac{1}{2}\{h,h\} +h^*_7 \bp\Lambda + h_7^*\{h,\Lambda\}\Big)\del(\langle\vep\lambda\rangle).
\end{equation}
Similarly to the gauge theory, we have the gauge freedom 
\begin{equation}
\del_{h_1}h^*_7 = \bp h_1 + \{h,h_1\}
\end{equation}
with $h^*_7\in\Omega^{0,2}(\PT,\O(1))$ and $h_1\in\Omega^{0,1}(\PT,\O(1))$. Writing 
$h_7^*$ in components: 
\begin{equation}h_7^*=h_{7,0\dal}^*\ov{e}^{0}\wedge\ov{e}^{\dal}+h_{7,\dal\dbe}^*\ov{e}^{\dal}\wedge\ov{e}^{\dbe}
\end{equation} 
we note that $h_{0\dal}$ has weight four while $h^*_{7,\dal\dbe}$ has weight three. As was the case for $\N=1$ sYM, to eliminate $h_7^*$ and $\Lambda$ from the action, it is enough to show that this gauge freedom can be used to set $h^*_{7,0\dal}$ to zero. For this, we pick a gauge such that $h_0=0$ \cite{Mason:2007ct}, so that we can write the gauge freedom as $\del_{h_1}h^*_{7,o\dal} = \bp_0 h_1$. From here we can follow the same analysis of section \ref{sec:N=1vecMultiplet} performing the expansion in weighted spherical harmonics. Since the weight of the gauge parameter $h_1$ is bigger then two, we follow the same steps as in the gauge theory to eliminate the $h^*_{7,0\dal}$ component. Consequently, we can integrate out $h^*_7$ and $\Lambda$ from the action. At the end, we are left with 
\begin{equation}
    S=\int D^3Z \, \tilde h\left(\bp h + \frac{1}{2}\{h,h\}\right)\del(\langle\vep\lambda\rangle)
\end{equation}

The twist of the non-linear graviton followed very closely the twist in the gauge theory, essentially just replacing the commutators by the Poisson bracket. 
We expect that the theory on spacetime to be a holomorphic Poisson BF theory \cite{Williams:2020fwd}, which is what we get by a naive reduction ignoring the compact nature of the twistor sphere. A proper reduction to spacetime is complicated by the Poisson brackets \cite{Sharma:2021pkl,Bittleston:2022nfr}, and we leave it for future work, along with the twist of the $\N=1$ matter fields.

\section{Holographic duals}\label{sec:Holography}

In this section we will use the framework of Chiral holography \cite{Sharma:2025ntb} to compute the bulk-duals to the minimal supersymmetric twist and chiral twist of $\N=4$ sYM discussed above. Chiral holography gives a construction of a dual holographic pair to self-dual $\N=4$ sYM from open-closed duality of the topological B-model. The supersymmetric twist of these dual pairs should then recover some examples of twisted holography \cite{Costello:2018zrm} and we will see that it is indeed the case. We compare our results with the work of \cite{Jarov:2025qhz} which performed this analysis for, among others, the chiral algebra twist, finding a localization to the twistor sphere instead of a plane on spacetime. We start with a brief review of the relevant aspects we will need, further details of the holographic setup can be found in \cite{Sharma:2025ntb,Costello:2018zrm}.

\subsection{Review on chiral holography}\label{ReviewChiralHolography}

In \cite{Sharma:2025ntb} the authors conjectured a duality between holomorphic Chern-Simons on supertwistor space, and the BCOV theory \cite{Bershadsky:1993cx} on the total space $X$ of a fibration $X=\mathcal{O}(-1)^{\oplus 4}\rightarrow \mathbb{PT}$ in the presence of a certain background field. This background field is sourced by a brane wrapping $\PT$ and the dual geometry is obtained by backreating this brane in $X$.



To describe this construction denote the homogeneous coordinates on $X$ as $(Z^A,W_i)$, with $Z^A = (\mu^{\dal},\lambda_\al)$ homogeneous coordinates on twistor space, and $W_i$ coordinates on the $\O(-1)$ fibers.
The string-field theory for the closed sector of the B-model is given by BCOV, or Kodaira-Spencer, theory \cite{Bershadsky:1993cx}. The B-model depends on the complex structure of target space, in our case the 7-fold $X$, which is Calabi-Yau with top-holomorphic form
\begin{equation}\label{TopForm}
\Omega=D^3Z\wedge d^4W\quad D^3Z=\vep_{ABCD}Z^AdZ^BdZ^CdZ^D\quad d^4W =\epsilon^{ijkl}dW_idW_jdW_kdW_l
\end{equation}
In this background the fields of BCOV are found inside the poly-vector fields:
\begin{equation}
    \Psi \in \bigoplus_{i,j}\text{PV}^{i,j}(X) \quad \text{PV}^{i,j}(X) = \Omega^{*,i}(X,\wedge^jTX).
\end{equation}
with $TX$ the holomorphic tangent to $X$. The top-holomorphic can be used to define the divergence operator $\p_\Omega:$PV$^{i,j}\rightarrow$PV$^{i,j+1}$ by 
\begin{equation}
    \p_{\Omega}\Psi \lrcorner \,\Omega = \p(\Psi\lrcorner\, \Omega),
\end{equation}
where $\p$ is the holomorphic exterior derivative. 
The fields of BCOV are then the divergence-free poly-vector fields
\begin{equation}
\p_{\Omega}\Psi=0.
\end{equation}
We will also need the Schouten-Nijenhuis bracket $[\Psi,\Psi']$, which can be defined as the failure of the divergent to obey the Leibniz rule:
\begin{equation}
    \p_{\Omega}(\Psi\wedge\Psi') = \p_\Omega\Psi \wedge \Psi' + (-1)^{\text{deg}\Psi}\Psi\wedge\p_\Omega \Psi' + [\Psi,\Psi'].
\end{equation}
The action for the BCOV theory is written as
\begin{equation}
    S=\int\Omega\wedge\Big( \frac{1}{2}\p^{-1}_\Omega \mathbf{\Psi}\ov{\p}\mathbf{\Psi} + \frac{1}{3!}\mathbf{\Psi}^3\Big)\lrcorner\,\Omega.
\end{equation}
The kinetic term posses a mild non-locality in $\p_{\Omega}^{-1}\Psi$ but it is well defined since the fields are divergence-free. This is the action for the whole BV complex of fields, and it only closes on the physical fields in complex dimension 3. By varying the action, we get the equations of motion:
\begin{equation}
    \bp\mathbf{\Psi} + \frac{1}{2}[\mathbf{\Psi},\mathbf{\Psi}] = 0,
\end{equation}
where we used that $\mathbf{\Psi}$ is divergent free to write $\p_\Omega(\mathbf{\Psi}\wedge \mathbf{\Psi}) = [\mathbf{\Psi},\mathbf{\Psi}]$.

For the open string sector we wrap $N$ D5-branes on $\PT\subset X$. The boundary conditions that preserve the B-model BRST symmetry only allow for fields on the brane that have anti-holomorphic form degrees along the brane, $d \overline{Z}^A$, and depend on the holomorphic normal directions $\frac{\p}{\p W_i}$. Identifying the latter as the fermionic $\eta_i\sim \frac{\p}{\p W_i}$ the theory on the brane is identified with the BV complex of holomorphic Chern-Simon on $\PT^{3|4}$ reviewed in section \ref{sec:SDSYM}. Chiral holography then posits that it can be can equivalently described by BCOV in the backreated geometry. To compute this backreaction we take into account the modification of the equations of motion of BCOV due to the presence of the brane. In the B-model, this is given by a disk amplitude with only one closed string insertion, which contributes to the action the term
\begin{equation}
    \int_{\mathbf{PT}}\p_{\Omega}^{-1}(\Psi\lrcorner \,\Omega).
\end{equation}
Where the integral is non-zero only for a top-form on $\PT$. This gives a delta-function source localized to the position of the brane to the equations of motion.

To find explicitly which fields are turned on by the backreaction we expand $\mathbf \Psi$ in terms of fields of definite vector degree:
\begin{equation}\label{PolyExpansion}
    \mathbf{\Psi} =\al + \be + \ga +\xi + \tilde \ga + \tilde \be + \tilde \al, 
\end{equation}
with $\al \in $ PV$^{*,0}(X)$, $\be \in $ PV$^{*,1}(X)$, $\ga \in $ PV$^{*,2}(X)$, $\xi \in $ PV$^{*,3}(X)$, $\tilde \ga \in $ PV$^{*,4}(X)$, $\tilde \be \in $ PV$^{*,5}(X)$ and $\tilde \al \in $ PV$^{*,6}(X)$\footnote{There is no PV$^{*,7}$ component since $\mathbf \Psi$ must be $\p_\Omega$ exact.}. The undeformed equations of motion in terms of this decomposition are:
\begin{align}
    &\bp \al +[\be,\al] = 0\\&\bp \be +\frac{1}{2}[\be,\be] + [ \ga,\al]= 0\\&\bp \ga +[\be,\ga] + [\xi,\al] = 0\\& \bp \xi +[\be,\xi] +[\tilde \ga,\al] + \frac{1}{2}[\ga,\ga]= 0\\&\bp \tilde\ga +[\be,\tilde \ga]  + [\tilde \be,\al] + [\ga,\xi]= 0\\& \bp \tilde\be +[\be,\tilde \be]  + [\tilde \al,\al] + [\ga,\tilde \ga] + \frac{1}{2}[\xi,\xi]= 0\\& \bp \tilde\al +[\be,\tilde \al]   + [\ga,\tilde \be] + [\xi,\tilde \ga]= 0.
\end{align}

Since $\PT$ is a complex 3-fold, the only term that gets a contribution from the source is $\xi$. The equation of motion in presence of the branes gets deformed to 
\begin{equation}
    \bp \xi  = N\del^4(W)d^4W,
\end{equation}
with solution 
\begin{equation}\label{Globalxi}
    \xi\lrcorner\, \Omega = N\frac{D^3\ov{W}d^4W}{||W||^8}.
\end{equation}
in the gauge $\overline{\p}^\dagger\xi=0$, where we picked some auxiliary hermitian metric. It will be important to us to have a local expression for $\xi$. 
On the affine patch such that $A^\al\lambda_\al\neq 0$, with $A^\al$ a reference spinor, we can write $\xi$ as 
\begin{equation}\label{localXI}
    \xi = \frac{N}{\langle A\lambda\rangle}\frac{D^3\ov{W}}{||W||^8}\vep^{\dal\dbe}A^\al\frac{\p}{\p\lambda{_\al}}\frac{\p}{\p\mu^{\dal}}\frac{\p}{\p\mu^{\dbe}}.
\end{equation}
This dependence on the reference spinor will be important for our analysis. Different choices of $A^\al$ can be interpreted as the same solution $\xi$ expressed locally on different patches where $\langle A\lambda\rangle\neq 0$. Any difference between the expressions on different patches must come from terms which are pure gauge\footnote{There could also have terms that are proportional to the vector field \begin{equation}
    V= w_i\frac{\p}{\p w_i} - Z^A\frac{\p}{\p Z^A},
\end{equation}
which vanishes projectively. However this will not be necessary for our analysis.}. The bulk dual to self-dual $\N=4$ sYM in twistor space is then BCOV theory on $\tilde X=X\setminus\mathbb{PT}$ with a background value for $\xi$ as given above.

\subsection{The bulk \texorpdfstring{$\mathfrak{psl}(4|4)$}{psl(4|4)} algebra}

The bulk BCOV theory should be thought of as a gravitational theory, where the supersymmetric twist is done by given a non-zero vev to the bosonic ghost corresponding to the twisting supersymmetry charge \cite{Costello:2016mgj}. The vev of this bulk field couples to the corresponding supersymmetry in the brane twisting it. One way to identify which bulk field should be given a vev is to find a realization, under the Schouten-Nijenhuis bracket, of the $\mathfrak{psl}(4|4)$ algebra in terms of bulk fields, similar to what was done in \cite{Costello:2018zrm,Costello:2016mgj}.

Looking at the $\mathfrak{psl}(4|4)$ generators in twistor space \eqref{FirstBasis}, we can identify the fields in the bulk corresponding to the supercharges 
\begin{align}
    &\lambda_\al W_i \quad\quad\quad\text{ordinary supercharges,}\\&\p_{\mu_{\dal}}\p_{W_i} \quad\quad\quad\text{conjugate ordinary supercharges,}\\&\p_{\lambda_{\al}}\p_{W_i} \quad\quad\quad\text{superconformal supercharges,}\\&\mu_{\dal} W_i \quad\quad\quad\text{conjugate superconformal supercharges.}\label{supcfchargebulk}
\end{align}
This can be easily checked using the Schouten-Nijenhuis bracket, for example 
\begin{equation}
    [\p_{\mu^{\dal}}\p_{W_i},W_j\lambda_\al] = -\del^i_j\lambda_\al\frac{\p}{\p\mu^{\dal}} = -\del^i_j P_{\al\dal},
\end{equation}
or 
\begin{align*}
    &[\lambda_\al W_i,\p_{\lambda_\be}\p_{W_j}] = \del_\al^\be W_i\frac{\p}{\p w_j} - \del_i^j\lambda_\al\frac{\p}{\p\lambda_\be} \\=&\del_\al^\be\Big(W_i\frac{\p}{\p W_j} - \frac{1}{4}\del_i^j W_k\frac{\p}{\p W_k}\Big) - \del_i^j\Big(\lambda_\al\frac{\p}{\p\lambda_\be}-\frac{1}{2}\del_\al^\be\lambda_\sigma\frac{\p}{\p\lambda_\sigma}\Big) + \frac{1}{2}\del_\al^\be\del_i^j\Big(\frac{1}{2}W_m\frac{\p}{\p W_m} - \lambda_\ga\frac{\p}{\p\lambda_\ga}\Big),
\end{align*}
which is the commutator (\ref{ComutatorQandS}). This gives a realization of the $\mathfrak{psl}(4|4)$ bulk algebra for $N=0$, that is, in the absence of the $D5$ brane. In the backreacted geometry we need to add $\mathcal{O}(N)$ corrections, such that the generators are holomorphic in the deformed geometry. Their expressions can be found in \cite{Sharma:2025ntb}, where the authors studied the deformed supergeometry. 

The fields we'll be interested in are $\lambda_\al W_i$, and $\mu_{\dal} W_i$, which correspond, respectively, to the action of $Q$ and $\tilde{S}$ on the boundary theory. Turning on these as background fields gives a superpotential to the B-model, localizing it to $\lambda=0=W$( respectively $\mu=0=W$) reproducing the localization of the gauge theory on twistor space, the reduced R-symmetry of the twisted sector.


\subsection{Localizing with a superpotential and twisted holography}


Analogously to the twist in the gauge theory, we parametrize the choice of supercharge by a spinor $\vep^\al$, turning on the following superpotential as a background field:
\begin{equation}
    \mathcal{W}_1 = \vep^\al\lambda_\al W_1.
\end{equation}
Without loss of generality, we have arbitrarily chosen the $w_1$ direction. This superpotential, together with the original background $\xi$ must obey the coupled background field equations of motion which can be read from further deforming BCOV by the superpotential
\begin{equation}\label{EOMsuperpotential}
    \bp \Psi_N + [\Psi_N,\Psi_N]+[\mathcal{W}_1,\Psi_N] = 0.
\end{equation}
$\Psi_N$ contains other background fields that might have to be turned on to ensure the equations of motion are satisfied. Turns out only one extra field has to be turned on as a background which is $\ga \in$ PV$^{*,2}$. Plugging this into \eqref{EOMsuperpotential} we find the following equations
\begin{equation}\label{twice_deformed}
    \bp \ga + [\mathcal{W}_1,\xi]=0\quad  [\mathcal{W}_1,\ga] = 0.
\end{equation}
We first look at the $[\mathcal{W}_1,\xi]$ term. Using the local expression (\ref{localXI}) we find 
\begin{equation}\label{commutator}
    [\mathcal{W}_1,\xi] = -N\frac{\langle A \vep\rangle}{\langle A \lambda\rangle}\frac{\ep_{ijkl}\ov{W}^id\ov{W}^jd\ov{W}^kd\ov{W}^l}{||W||^8}W_1\vep^{\dal\dbe}\p_{\mu^{\dal}}\p_{\mu^{\dbe}}.
\end{equation}
There is an important feature of this commutator: recall that we can choose $A^\al$ arbitrarily, as long as $\langle A \lambda\rangle \neq 0$ on the patch. In particular, on the patch that excludes the point $\langle \lambda\vep\rangle =0$ we can choose $A^{\al} = \vep^{\al}$, so that the commutator is zero on this patch. Different choices of reference spinor give different polyvectors when treated non-projectively on $\mathbb{C}^8$. However, projectively, two polyvectors with different reference spinors must descend, up to gauge, to the same polyvector on $\tilde{X}$ on overlaps where both are defined. From the argument above, if we choose any patch that does not contain the point $\langle \lambda\vep\rangle=0$, then this polyvector vanishes everywhere on this patch. Hence, there is only one point where this polyvector does not vanish: precisely on $\langle\lambda\vep\rangle=0$. This commutator is effectively supported on a delta function, providing a localization on $\langle\lambda\vep\rangle=0$ in the bulk theory. 
Hence, we write 
\begin{equation}
    [\mathcal{W}_1, \xi] = -N\del(\langle\lambda\vep\rangle)\frac{\ep_{ijkl}}{||W||^8}W_1\ov{W}^id\ov{W}^jd\ov{W}^kd\ov{W}^l\vep^{\dal\dbe}\p_{\mu^{\dal}}\p_{\mu^{\dbe}}.
\end{equation}
It is then easy to check that  
\begin{equation}\label{gamma}
    \ga = \frac{N}{3} \del(\langle\lambda\vep\rangle)\ep_{1ijk}\frac{\ov{W}_id\ov{W}_jd\ov{W}_k}{||W||^6}\vep^{\dal\dbe}\p_{\mu^{\dal}}\p_{\mu^{\dbe}}.
\end{equation}
is indeed a solution of \eqref{twice_deformed}. The commutator $[\mathcal{W}_1,\ga]$ is trivially zero, since the superpotential doesn't depend on $\mu^{\dal}$.

This result is in agreement with \cite{Costello:2016mgj,Costello:2018zrm}, where the authors conjectured that the holographic dual of the holomorphic twist of $\N=4$ is the B-model on $\mathbb C^5\verb|\|\mathbb C^2$ with a specific polyvector turned on, given by 
\begin{equation}
    F = N\ep_{ijk} \frac{\ov{w}_id\ov{w}_jd\ov{w}_k}{||w||^6}\p_{z_1}\p_{z_2}\in \text{PV}^{2,2}(\mathbb C^5\verb|\|\mathbb C^2).
\end{equation}
The expression in $(\ref{gamma})$ gives the 7-fold analog of this polyvector. When restricted to $\langle\lambda\vep\rangle=0$ and $w_1 = 0$, it precisely reproduces $F$. 

To compute the bulk dual to the chiral algebra twist we will follow a very similar procedure, finding the same deformed conifold of twisted holography \cite{Costello:2018zrm}. We need to turn on a superpotential corresponding to the superconformal charge $\tilde S$ on the gauge theory side. From equation (\ref{supcfchargebulk}), we know that the superpotential should have the form $\mu^{\dal}w_i$. As before, parametrize this choice by a constant spinor $\epb^{\dal}$ writing it as
\begin{equation}
    \mathcal{W}_2 = \epb_{\dal}\mu^{\dal} W_2. 
\end{equation}
The first thing we should notice is that $[\xi,\mathcal{W}_2]\neq 0$. This means that this superpotential will change the solution for the bulk fields on $X$.  
From the degree of $\ga$ and $\xi$ we see that we should turn on two polyvectors, $\be\in \text{PV}^{*,1}$ and $\ov \ga\in \text{PV}^{*,2}$, obeying the equations of motions:
\begin{align}
    &[\be,\mathcal{W}_1]+[\be,\mathcal{W}_2]=0\\&\bp\be +[\be,\be]+[\mathcal{W}_2,\ga] + [\mathcal{W}_1,\ov \ga]+[\mathcal{W}_2,\ov \ga] = 0\\&\bp \ov\ga + [\mathcal{W}_2,\xi] + [\be,\ga] + [\be,\ov\ga]=0\\&[\xi,\be]=0 \\& [\xi,\ov\ga]=0.
\end{align}
A short computation gives the commutator
\begin{equation}\label{beta}
    [\mathcal{W}_2,\ga] = -\frac{N}{3}\del(\langle\lambda\vep\rangle)\ep_{1ijk}\frac{\ov{W}_id\ov{W}_j d\ov{W}_k}{||W||^6}W_2\p_z
\end{equation}
with
\begin{equation}
    \p_z = \epb_{\dot 1}\frac{\p}{\p z^{\dot 2}} - \epb_{\dot 2}\frac{\p}{\p z^{\dot 1}}.
\end{equation}
This expression shows how $\epb^{\dal}$ parametrizes the choice of the chiral algebra plane, with $z$ being the holomorphic coordinate.
Similarly, using the expression \eqref{localXI} we find 
\begin{equation}
    [\mathcal{W}_2,\xi] = -\frac{N}{\langle A\lambda\rangle}\frac{D^3\ov{W}}{||W||^8}A^\al\frac{\p}{\p\lambda^\al}\p_z
\end{equation}
To find the solution for $\ov \ga$, we first find the solution for $\bp\ov\ga +[\mathcal{W}_2,\xi]=0$. The solution is analogous to \eqref{gamma}:
\begin{equation}
    \ov \ga = \frac{N}{3\langle A\lambda\rangle} \ep_{2ijk}\frac{\ov{W}_id\ov{W}_jd\ov{W}_k}{||W||^6}A^\al\frac{\p}{\p\lambda^\al}\p_z.
\end{equation}
Now, we can compute the commutators of $\ov \ga$. It is easy to see that $\p_z\mathcal{W}_2 \propto \epb_{\dal}\epb^{\dal} = 0$, hence $[\ov\ga,\mathcal{W}_2]=0$. For $\mathcal{W}_1$ we find
\begin{equation}
    [\mathcal{W}_1,\ov\ga] = -\frac{\langle A\vep\rangle}{3\langle A\lambda\rangle}\ep_{2ijk}\frac{\ov{W}_id\ov{W}_jd\ov{W}_k}{||W||^6}W_1\p_z.
\end{equation}
Using the same argument of equation \eqref{commutator}, we can replace $\langle A\vep\rangle/\langle A\lambda\rangle$ by a delta function. Note the similarity of the commutators of $[\ga,\mathcal{W}_2]$ and $[\ov\ga,\mathcal{W}_1]$. This gives the following solution for $\be$:
\begin{equation}
    \beta = \frac{N}{3}\del(\langle\lambda\vep\rangle)\frac{\ep_{12ij}\ov{W}^id\ov{W}^j}{||W||^4}\p_{z}.
\end{equation}
Again, we find $\bp\be+[\mathcal{W}_2,\ga] + [\mathcal{W}_1,\ov\ga]=0$. One can check that all other commutators vanish. 

When restricted to the the locus $w_2=w_1=0$ and $\langle\lambda\vep\rangle=0=\epb_{\dal}\mu^{\dal}$, the Beltrami differential is holomorphic and gives an integrable deformation of the complex structure. The bulk spacetime is then $\mathbb{C}^3$ with a deformed complex structure given by this Beltrami reproducing the geometry found in \cite{Costello:2018zrm} for the bulk dual to the chiral algebra of $\N=4$ sYM. 

\section{Discussion}

We have shown that the minimal supersymmetric twist of self-dual Yang-Mills in twistor space coincides perturbatively with the analogous supersymmetric twist of Yang-Mills on spacetime. Since the this twist on twistor spaces localizes to a holomorphic theory on spacetime, any further consistent supersymmetric twist on twistor space will coincide with its spacetime counterpart. We hope that this will serve to further motivate a deeper study of BPS sectors on twistor space, and their holographic duals.

We also briefly studied the holographic duals to the minimal, and chiral algebra twists of $\N=4$ sYM in the framework of chiral holography, showing that both reproduced known results in the twisted holography literature. Another work that addresses, among others, the chiral algebra twist and its dual in the same framework is \cite{Jarov:2025qhz}. Contrary to this work, we find that the chiral algebra localizes, as usual, to a plane inside spacetime, while  \cite{Jarov:2025qhz} found that the chiral plane is localized to the twistor sphere. This can be traced back to the choice of twisting supercharges in that work, which are unusual from a spacetime point of view. Nevertheless, the same geometry and holographic duality are recovered in the end. We leave a better understanding of the relation between these two ways of performing this particular twist for future work.

The twisted theories in twistor space end up being topological-holomorphic. That means we can move local operators along the topological direction and thus expect that there are no short-distance singularities. But operator products in holomorphic BF theory in four dimensions can have short-distance singularities. It was shown in \cite{Bomans:2025klo} how an analogous puzzle is resolved in the chiral algebra twist of holomorphic BF theories. It would be interesting to do a similar analysis in our setup.

Another puzzle comes from using an axial gauge in twistor space. This gauge choice kills all the tree-point interactions, rendering the theory effectively free. Nevertheless, axial gauge has found great success in the computation of tree-level amplitudes and of integrands \cite{Bullimore:2010pj,Mason:2010yk,Adamo:2011cb,Adamo:2011pv}, as well as in the study of correlation functions \cite{Koster:2016fna,Koster:2016loo,Koster:2016ebi,Caron-Huot:2023wdh}. The axial gauge seems to preclude any loop corrections that are known to be present in the BPS sector, and we should revisit the use of this gauge.

\subsection*{Acknowledgments}
We thank Atul Sharma for comments on an early version of this work. This study was financed, in part, by the São Paulo Research Foundation (FAPESP), Brasil, Process Number 2024/12765-6, CNPq under grant 303408/2024-3, by CAPES under grant 88882.461730/2019-01.
\appendix

\section{BV action of $\N=2,4$ self-dual sYM on twistor space}\label{app:BVaction}

In this appendix we will find the BV action for the $\N=2$ and $\N=4$ self-dual theories on twistor space. Self-dual $\N=2$ is described in twistor space as a BF theory in $\mathbb{PT}^{3|2}$. As we explained on section \ref{sec:BVactionN=1}, the BV theory for holomorphic BF is
\begin{equation}\label{BVofBF}
    S = \int D^{3|2}\mathcal{Z}\,(\mathbf B\ov{\mathbf D} \mathbf A +\mathbf A^* \ov{\mathbf D}\mathbf c + \mathbf B^*\left(\ov{\mathbf D}\mathbf\xi + [\mathbf c,\mathbf B]\right) + \frac{1}{2}\mathbf c^*[\mathbf c, \mathbf c] + \mathbf \xi^*[\mathbf c,\mathbf \xi]),
\end{equation}
where $\mathbf c, \mathbf\xi \in \Omega^{0,0}(\mathbb{PT}^{3|2})$, $\mathbf A, \mathbf B\in \Omega^{0,1}(\mathbb{PT}^{3|2})$, $\mathbf A^*, \mathbf B^*\in \Omega^{0,2}(\mathbb{PT}^{3|2})$ and $\mathbf c^*, \mathbf\xi^*\in \Omega^{0,3}(\mathbb{PT}^{3|2})$. All components are functions of $\eta_1,\eta_2$ and $Z^A$. Doing the expansion in the fermionic coordinates of section \ref{sec:SDSYM}
\begin{equation}
\begin{aligned}
    &\mathbf{B} = \ov{\phi} + \frac{1}{2}\ep^{ij}\eta_i\chi_j + \ep^{ij}\eta_i\eta_j b,\quad \mathbf{A} = a + \eta_i\psi^i + \frac{1}{2}\ep^{ij}\eta_i\eta_j \phi,\\&\mathbf{A^*} = \phi^*+\ep^{ij}\psi^*_i\eta_j+\frac{1}{2}\ep^{ij}\eta_i\eta_j a^*, \quad \mathbf{B^*} = b^*+\chi^{ i*}\eta_i + \frac{1}{2}\ep^{ij}\eta_i\eta_j\ov{\phi}^*,\\&\mathbf{c} = c+\Gamma^i\eta_i+\frac{1}{2}\ep^{ij}\eta_i\eta_j \sigma, \quad \mathbf{\xi} = \ov{\sigma}+\Lambda^i\eta_i+\frac{1}{2}\ep^{ij}\eta_i\eta_j \xi,\\&\mathbf{c^*} = \sigma^*+\Gamma^{i*}\eta_i+\frac{1}{2}\ep^{ij}\eta_i\eta_j c^*, \quad \mathbf{\xi^*} = \xi^*+\Lambda^{i*}\eta_i+\frac{1}{2}\ep^{ij}\eta_i\eta_j \ov{\sigma}^*.
\end{aligned}
\end{equation}
This yields the action 
\begin{equation}\label{BVN=2}
\begin{aligned}
    S_{\text{BV}}&=S_\text{cl} +\int_{\mathbb{PT}} a^*\ov{D}c +\psi^{*_i}\left(\ov{D}\Gamma^i+[c,\psi^i]\right)+\phi^*\left(\ov{D}\sigma+[\Gamma_i,\psi^i]+[c,\phi]\right)\\&+\ov{\phi}^*\left(\ov{D}\ov{\sigma}+[c,\ov{\phi}]\right)+\chi^{i*}\left(\ov{D}\Lambda_i -[\psi_i,\ov{\sigma}]+[\Gamma_i,\ov{\phi}]\right) +b^*\Big(\ov{D}\xi +[\Gamma^i,\chi_i]\\& - [\Lambda_i,\psi^i]-[{\phi},\ov{\sigma}]+[\ov{\phi},\sigma]\Big)+\frac{1}{2}c^*[c,c] + \Gamma_{i}^*[c,\Gamma^i] + \sigma^*[c,\sigma]\\&+\frac{1}{2}\sigma^*[\Gamma^i,\Gamma_i]+\ov{\sigma}[c,\ov{\sigma}]+\Lambda^{i*}[c,\Lambda_i]+\xi^*[c,\xi] + \xi^*[\Gamma^i,\Lambda_i],
\end{aligned}
\end{equation}
where we omited the holomorphic measure for clarity. Here $\Gamma_i,\Lambda_i,\sigma$ and $\ov{\sigma}$ are ghosts for gauge transformations. The terms that are linear in the antifields multiply the gauge transformation of each physical fields. The terms that are linear the antifields from the ghosts encode the non-trivial structure components of the gauge algebra.

Now, lets turn to the construction of the BV action of self dual $\N=4$ in twistor space. This is described by holomorphic Chern-Simons in twistor space (\ref{HoloCSTwistor}). To find the BV action, lets first write the BV action of holomorphic Chern-Simons. Lets do the expansion in fields with varying ghost number:
\begin{equation}
    \mathcal{A} = \mathbf{c}+\mathbf{a}+\mathbf{a}^* + \mathbf{c}^*,
\end{equation}
where $\mathbf{c}\in\Omega^{0,0}(\mathbb{PT}^{3|4})$, $\mathbf{a}\in\Omega^{0,1}(\mathbb{PT}^{3|4})$, $\mathbf{a^*}\in\Omega^{0,2}(\mathbb{PT}^{3|4})$ and $\mathbf{c^*}\in\Omega^{0,3}(\mathbb{PT}^{3|4})$. All fields have ghost number $g = 1-p$, where $p$ is the form degree. The action is
\begin{equation}
    S=\frac{1}{2}\int D^{3|4}\mathcal{Z}(\mathbf{a}\bp\mathbf{a}+\frac{2}{3}\mathbf{a}^3+a^*\ov{D}c + \frac{1}{2}\mathbf{c}^*[c,c]).
\end{equation}
Of course, the first two terms will give the classical action (\ref{n=4action}). Here we will focus on the last two terms. We need to perform the fermionic expansion:
\begin{equation}
\begin{aligned}
    \\&\mathbf{a} = a +\psi^i\eta_i+\frac{1}{2}\phi^{ij}\eta_i\eta_j + \frac{1}{3!}\ep^{ijkl}\chi_i\eta_j\eta_k\eta_l+\frac{1}{4!}b\eta^4,\\&\mathbf{a^*} = b^*+\chi^{i*}\eta_i + \frac{1}{2}\phi^{ij*}\eta_i\eta_j+ \frac{1}{3!}\ep^{ijkl}\psi^*_i\eta_j\eta_k\eta_l+\frac{1}{4!}a^*\eta^4,\\&\mathbf{c} = c +\Gamma^i\eta_i+\frac{1}{2}\sigma^{ij}\eta_i\eta_j + \frac{1}{3!}\ep^{ijkl}\Lambda_i\eta_j\eta_k\eta_l+\frac{1}{4!}\xi\eta^4,\\&\mathbf{c^*} = \xi^*+\Lambda^{i*}\eta_i + \frac{1}{2}\sigma^{ij*}\eta_i\eta_j+ \frac{1}{3!}\ep^{ijkl}\Gamma^*_i\eta_j\eta_k\eta_l+\frac{1}{4!}c^*\eta^4.
\end{aligned}
\end{equation}
Where we have introduced one ghost for each matter field. Then, the BV action becomes 
\begin{equation}
\begin{aligned}\label{BVN=4}
    S_{\text{BV}} &= S_{\text{cl}} + \int_{\mathbb{PT}}a^*\ov{D}c + \psi_i^*\left(\ov{D}\Gamma^i+ [c,\psi^i]\right) + \phi_{ij}^*\Big(\ov{D}\sigma^{ij} +[\psi^i,\Gamma_i] +[c,\phi^{ij}]\Big)\\&+\chi^{i*}\Big(\ov{D}\Lambda_i-[\psi^j,\sigma_{ij}]+ [\Gamma^j,\phi_{ij}] + [c,\chi_i]\Big)+b^*\Big(\ov{D}\xi - [\psi^i,\Lambda_i]\\&-\ep^{ijkl}[\phi_{ij},\sigma_{kl}] + [\chi_i,\Gamma^i]+[c,b]\Big)+ \xi^*\left([c,\xi]+[\Gamma^i,\Lambda_i]+\ep_{ijkl}[\sigma^{ij},\sigma^{kl}]\right)\\&+\sigma^{*ij}\left([c,\sigma_{ij}]+\frac{1}{2}[\Gamma_i,\Gamma_j]\right)+\Lambda^{i*}\left([c,\Lambda_i]+[\Gamma^i,\sigma_{ij}]\right)+\frac{1}{2}c^*[c,c] + \Gamma^*_i[c,\Gamma^i],
\end{aligned}
\end{equation}
where we again omitted the holomorphic measure.

\section{Reduction to spacetime of twisted $\N=4$ actions}\label{app:ReductionToSpacetime}

In this section we will show the details of how to perform the reduction to spacetime of the actions of the twisted theories that we found in the main text. Below we omit the holomorphic measure for clarity.

\subsection{Holomorphic twist of $\N=4$}\label{subapp:ReducHoloN=4}
Lets first perform this for $\N=4$, following the results from section \ref{N=4Holotwist}.  

First, we want to localize the action (\ref{TwistedN=4}) to a point. For this, we will use the gauge freedom of $\del_\xi$ for $\Lambda_1$ and the gauge freedom of $\sigma^{1I}$ for $\Gamma^I$ to set these fields to zero everywhere, except at $\langle\lambda\vep\rangle = 0$. Then we perform field redefinitions very similar to (\ref{fieldRedefinitionsTwo}): 
\begin{equation}
        \chi^*_1\vep^\al\lambda_\al b \rightarrow \chi^*_1\vep^\al\lambda_\al b -b\ov{Da}\quad \psi_I^*\vep^\al\lambda_\al \phi_2 \rightarrow \psi_I^*\vep^\al\lambda_\al \phi - {\Phi}^I\,\ov{D}\Phi_I.
\end{equation}
The action localized to $\langle\lambda\vep\rangle =0$ is 
\begin{multline}\label{TwistedN=4}
    S_{\text{cl,}\N=4} = \int\Big[ b\ov{D}a + \Phi^{I}\ov{D}\Phi_{I}+ \chi^{1*}\left(\vep^\al \lambda_\al b +\ov{D}\Lambda_1 - [\Gamma^I,\phi_{I}]\right) \\+\psi^*_I\left(\ov{D}\Gamma^I + \vep^\al\lambda_\al \Phi^{I}\right)\Big]\del(\langle\lambda\vep\rangle).
\end{multline}
From here we follow the same steps as in the $\N=2$ case. First, we use the gauge freedom of $\del_{\psi^1}\chi^{1*} = \ov{D}\psi^1$ to set $\chi^{1*}_{0\dal}$ to zero. Since $\psi^{*I}\in\mathcal{O}(-3)$, we can directly use the results from $\N=2$ to conclude the gauge freedom of $\del_{\chi_I}\psi^{*I} =\ov{D}\chi_I$ can be used to eliminate all modes from the spherical harmonics expansion of $\psi^*_{I,0\dal}$ that depend on $\lambda$ and $\hat\lambda$. That is, we find $\psi^*_{I,0\dal} = \be_{\dal}^I$. 
Integrating out $\psi^*_{I,\dal\dbe}$ we find the constraint:
\begin{equation}
    \bp_0\Gamma^I\Big|_{\lambda_\al=\vep_\al} = 0,
\end{equation}
which has the solution
\begin{equation}
    \Gamma^I = \frac{\langle\vep\hat\lambda\rangle}{\langle\lambda\hat\lambda\rangle}\ga^I(x).
\end{equation}
Finally, by integrating out $\chi^{1*}_{\dal\dbe}$ we find
\begin{equation}
    \bp_0\Lambda^1\Big|_{\lambda_\al=\vep_\al} = [\Gamma^I,\Phi_I]\Big|_{\lambda_\al=\vep_\al} = \frac{\langle\vep\hat\lambda\rangle}{\langle\lambda\hat\lambda\rangle}[\ga^I,\Phi_{I,0}]\Big|_{\lambda_\al=\vep_\al}.
\end{equation}
Again the only solution is that $\Phi_{I,0}$ and $\Lambda^1$ are constants and can be discarded. Because of (\ref{WoodGaugeScalar}), this fully eliminated the scalars kinetic term. We are left with the action
\begin{equation}
    S_{\text{cl,}\N=4} = \int \del(\langle\lambda\vep\rangle)\left[b\ov{D}a + \frac{\langle\vep\hat\lambda\rangle}{\langle\lambda\hat\lambda\rangle}\be^{\dal}_I\ov{D}_{\dal}\ga^I\right].
\end{equation}
Integrating along $\CP^1$ using the delta function we find the action for the holomorphic $BF$ theory with three $\be\ga$ systems:
\begin{equation}
    S_{\text{cl,}\N=4} = \int b\ov{D}a + \be_I\ov{D}\ga^I,
\end{equation}
with 
\begin{equation}
    \be_I = \be_{I,\dal}d\ov{z}^{\dal}dz^{\dbe}dz_{\dbe} \quad \ov{D}=\vep^\ga d\ov{z}^{\dga}D_{\ga\dga}.
\end{equation}
\subsection{Chiral algebra twist of $\N=4$}\label{subapp:ReducChiralN=4}
As it was the case for the chiral algebra twist of the $\N=2$ multiplets, we can further localize the action in twistor space. For the $\Gamma^b$ fields, we note that they transform in the following way under the gauge transformations of $\sigma^{ab}$: 
\begin{equation}
    \del\Gamma^b = \vep^\al\lambda_\al\sigma^{1b}+\epb^{\dal}\mu_{\dal}\sigma^{2b}.
\end{equation}
Hence, we can use this gauge freedom to set $\Gamma^b$ to zero everywhere, except at $\langle\lambda\vep\rangle=0= \epb_{\dal}\mu^{\dal} $. Then integrate out $\sigma^{12^*}$, which yields the same constrain as (\ref{Ghostconstraint}). This means that we can again use the gauge freedom (\ref{GaugeXiCiralTwistN=2}) to localize the field $\Lambda_a$. Now perform the field redefinitions:
\begin{equation}\label{FirstFieldRedCHiralN=4}
    \chi^{*1}\vep^\al\lambda_\al \rightarrow \chi^{*1}\vep^\al\lambda_\al - \ov{D}a\quad \chi^{*1}\epb^{\dal}\mu_{\dal} \rightarrow \chi^{*2}\epb^{\dal}\mu_{\dal} - \ov{D}a,
\end{equation}
to localize the $b\ov{D}a$ term, as well as 
\begin{equation}\label{FieldRedScalarsCHiralN=4}
    \psi^*_3 \rightarrow \psi^*_3 + \frac{1}{\langle\vep\lambda\rangle}\ov{D}\phi^{24} + \frac{1}{[\epb\mu]}\ov{D}\phi^{14}\quad \psi^*_4 \rightarrow \psi^*_4 + \frac{1}{\langle\vep\lambda\rangle}\ov{D}\phi^{23} + \frac{1}{[\epb\mu]}\ov{D}\phi^{13}
\end{equation}
With these field redefinitions we have localized the action to the point $\langle\lambda\vep\rangle =0=\epb_{\dal}\mu^{\dal}$. One might worry that the field redefinitions (\ref{FieldRedScalarsCHiralN=4}) can introduce cross terms between the scalars, however, one can check that these terms cancel out. Hence, we are left with the action 
\begin{equation}
\begin{aligned}
    S_{\text{chiral }\N=4}=&\int \Bigg[b\ov{D}a+ \frac{1}{2}\ep_{ab}\ep_{aa'}\phi^{bb'}\ov{D}\phi^{a'b'} +\chi^{*1}\left(\ov{D}\Lambda_1 +[\Gamma^b,\phi_{1b}]\right)+\chi^{*2}\left(\ov{D}\Lambda_2+[\Gamma^b,\phi_{2b}]\right)\\& + \frac{\ep_{bb'}}{2}\sigma^{12*}  [\Gamma^b,\Gamma^{b'}] +\psi^*_b\ov{D}\Gamma^b \Bigg]\del(\langle\lambda\vep\rangle)\del\left(\epb^{\dal}\mu_{\dal}\right)
\end{aligned}
\end{equation}
From here, the reduction to the plane is completely analogous to the $\N=2$ and $\N=4$ cases done in sections \ref{subsec:N=2twist} and \ref{subapp:ReducHoloN=4}. At the end, we obtain the action (\ref{ChiralAlgebraN=4Action}), as expected.

\bibliographystyle{JHEP}
\bibliography{main}

@article{Adamo:2011pv,
    author = "Adamo, Tim and Bullimore, Mathew and Mason, Lionel and Skinner, David",
    title = "{Scattering Amplitudes and Wilson Loops in Twistor Space}",
    eprint = "1104.2890",
    archivePrefix = "arXiv",
    primaryClass = "hep-th",
    doi = "10.1088/1751-8113/44/45/454008",
    journal = "J. Phys. A",
    volume = "44",
    pages = "454008",
    year = "2011"
}

@article{Adamo:2011cb,
    author = "Adamo, Tim and Mason, Lionel",
    title = "{MHV diagrams in twistor space and the twistor action}",
    eprint = "1103.1352",
    archivePrefix = "arXiv",
    primaryClass = "hep-th",
    doi = "10.1103/PhysRevD.86.065019",
    journal = "Phys. Rev. D",
    volume = "86",
    pages = "065019",
    year = "2012"
}

@article{Koster:2016fna,
    author = "Koster, Laura and Mitev, Vladimir and Staudacher, Matthias and Wilhelm, Matthias",
    title = "{On Form Factors and Correlation Functions in Twistor Space}",
    eprint = "1611.08599",
    archivePrefix = "arXiv",
    primaryClass = "hep-th",
    reportNumber = "HU-MATHEMATIK-2016-20, HU-EP-16-40, MITP-16-123",
    doi = "10.1007/JHEP03(2017)131",
    journal = "JHEP",
    volume = "03",
    pages = "131",
    year = "2017"
}

@article{Koster:2016loo,
    author = "Koster, Laura and Mitev, Vladimir and Staudacher, Matthias and Wilhelm, Matthias",
    title = "{All tree-level MHV form factors in $ \mathcal{N} $ = 4 SYM from twistor space}",
    eprint = "1604.00012",
    archivePrefix = "arXiv",
    primaryClass = "hep-th",
    reportNumber = "HU-MATHEMATIK-2016-06, HU-EP-16-10, MITP-16-027",
    doi = "10.1007/JHEP06(2016)162",
    journal = "JHEP",
    volume = "06",
    pages = "162",
    year = "2016"
}

@article{Koster:2016ebi,
    author = "Koster, Laura and Mitev, Vladimir and Staudacher, Matthias and Wilhelm, Matthias",
    title = "{Composite Operators in the Twistor Formulation of N=4 Supersymmetric Yang-Mills Theory}",
    eprint = "1603.04471",
    archivePrefix = "arXiv",
    primaryClass = "hep-th",
    reportNumber = "HU-MATHEMATIK-16-05, HU-EP-16-09, MITP-16-024",
    doi = "10.1103/PhysRevLett.117.011601",
    journal = "Phys. Rev. Lett.",
    volume = "117",
    number = "1",
    pages = "011601",
    year = "2016"
}

@article{Mason:2010yk,
    author = "Mason, L. J. and Skinner, David",
    title = "{The Complete Planar S-matrix of N=4 SYM as a Wilson Loop in Twistor Space}",
    eprint = "1009.2225",
    archivePrefix = "arXiv",
    primaryClass = "hep-th",
    doi = "10.1007/JHEP12(2010)018",
    journal = "JHEP",
    volume = "12",
    pages = "018",
    year = "2010"
}

@article{Bullimore:2010pj,
    author = "Bullimore, Mathew and Mason, L. J. and Skinner, David",
    title = "{MHV Diagrams in Momentum Twistor Space}",
    eprint = "1009.1854",
    archivePrefix = "arXiv",
    primaryClass = "hep-th",
    doi = "10.1007/JHEP12(2010)032",
    journal = "JHEP",
    volume = "12",
    pages = "032",
    year = "2010"
}

@article{Mason:2007ct,
    author = "Mason, L. J. and Wolf, Martin",
    title = "{Twistor Actions for Self-Dual Supergravities}",
    eprint = "0706.1941",
    archivePrefix = "arXiv",
    primaryClass = "hep-th",
    reportNumber = "IMPERIAL-TP-MW-02-07",
    doi = "10.1007/s00220-009-0732-5",
    journal = "Commun. Math. Phys.",
    volume = "288",
    pages = "97--123",
    year = "2009"
}

@article{Wolf:2007tx,
    author = "Wolf, Martin",
    title = "{Self-Dual Supergravity and Twistor Theory}",
    eprint = "0705.1422",
    archivePrefix = "arXiv",
    primaryClass = "hep-th",
    doi = "10.1088/0264-9381/24/24/010",
    journal = "Class. Quant. Grav.",
    volume = "24",
    pages = "6287--6328",
    year = "2007"
}

@article{Williams:2020fwd,
    author = "Williams, Brian R. and Elliott, Chris",
    title = "{Holomorphic Poisson Field Theories}",
    eprint = "2008.02302",
    archivePrefix = "arXiv",
    primaryClass = "math-ph",
    doi = "10.21136/HS.2021.08",
    journal = "Higher Struct.",
    volume = "5",
    number = "1",
    pages = "282--309",
    year = "2021"
}

@article{Caron-Huot:2023wdh,
    author = {Caron-Huot, Simon and Coronado, Frank and M{\"u}hlmann, Beatrix},
    title = "{Determinants in self-dual $ \mathcal{N} $ = 4 SYM and twistor space}",
    eprint = "2304.12341",
    archivePrefix = "arXiv",
    primaryClass = "hep-th",
    doi = "10.1007/JHEP08(2023)008",
    journal = "JHEP",
    volume = "08",
    pages = "008",
    year = "2023"
}

@article{Bomans:2025klo,
    author = "Bomans, Pieter and Garner, Niklas and Williams, Brian R. and Wu, Jingxiang",
    title = "{Unravelling the Holomorphic Twist II: Anomalies and Extended Supersymmetry}",
    eprint = "2509.16737",
    archivePrefix = "arXiv",
    primaryClass = "hep-th",
    month = "9",
    year = "2025"
}

@article{Garner:2022its,
    author = "Garner, Niklas and Paquette, Natalie M.",
    title = "{Mathematics of String Dualities}",
    eprint = "2204.01914",
    archivePrefix = "arXiv",
    primaryClass = "hep-th",
    doi = "10.22323/1.403.0007",
    journal = "PoS",
    volume = "TASI2021",
    pages = "007",
    year = "2023"
}

@article{Eager:2018dsx,
    author = "Eager, Richard and Saberi, Ingmar and Walcher, Johannes",
    title = "{Nilpotence varieties}",
    eprint = "1807.03766",
    archivePrefix = "arXiv",
    primaryClass = "hep-th",
    doi = "10.1007/s00023-020-01007-y",
    journal = "Annales Henri Poincare",
    volume = "22",
    number = "4",
    pages = "1319--1376",
    year = "2021"
}

@article{Saberi:2021weg,
    author = "Saberi, Ingmar and Williams, Brian R.",
    title = "{Twisting pure spinor superfields, with applications to supergravity}",
    eprint = "2106.15639",
    archivePrefix = "arXiv",
    primaryClass = "math-ph",
    doi = "10.4310/PAMQ.2024.v20.n2.a2",
    journal = "Pure Appl. Math. Quart.",
    volume = "20",
    number = "2",
    pages = "645--701",
    year = "2024"
}

@article{Closset:2013vra,
    author = "Closset, Cyril and Dumitrescu, Thomas T. and Festuccia, Guido and Komargodski, Zohar",
    title = "{The Geometry of Supersymmetric Partition Functions}",
    eprint = "1309.5876",
    archivePrefix = "arXiv",
    primaryClass = "hep-th",
    reportNumber = "WIS-09-13-SEP-DPPA",
    doi = "10.1007/JHEP01(2014)124",
    journal = "JHEP",
    volume = "01",
    pages = "124",
    year = "2014"
}

@misc{costello2013notessupersymmetricholomorphicfield,
      title={Notes on supersymmetric and holomorphic field theories in dimensions 2 and 4}, 
      author={Kevin J. Costello},
      year={2013},
      eprint={1111.4234},
      archivePrefix={arXiv},
      primaryClass={math.QA},
      url={https://arxiv.org/abs/1111.4234}, 
}

@article{Losev:1996up,
    author = "Losev, Andrei and Moore, Gregory W. and Nekrasov, Nikita and Shatashvili, Samson",
    title = "{Chiral Lagrangians, anomalies, supersymmetry, and holomorphy}",
    eprint = "hep-th/9606082",
    archivePrefix = "arXiv",
    reportNumber = "PUPT-1627, ITEP-TH-18-96, YCTP-P10-96",
    doi = "10.1016/S0550-3213(96)00612-8",
    journal = "Nucl. Phys. B",
    volume = "484",
    pages = "196--222",
    year = "1997"
}

@article{Johansen:1994aw,
    author = "Johansen, A.",
    title = "{Twisting of $N=1$ SUSY gauge theories and heterotic topological theories}",
    eprint = "hep-th/9403017",
    archivePrefix = "arXiv",
    reportNumber = "FERMILAB-PUB-93-062-TA",
    doi = "10.1142/S0217751X9500200X",
    journal = "Int. J. Mod. Phys. A",
    volume = "10",
    pages = "4325--4358",
    year = "1995"
}

@article{Baulieu:2004pv,
    author = "Baulieu, Laurent and Tanzini, Alessandro",
    title = "{Topological symmetry of forms, N=1 supersymmetry and S-duality on special manifolds}",
    eprint = "hep-th/0412014",
    archivePrefix = "arXiv",
    reportNumber = "PAR-LPTHE-P04-32, SISSA-94-2004-FM",
    doi = "10.1016/j.geomphys.2005.12.006",
    journal = "J. Geom. Phys.",
    volume = "56",
    pages = "2379--2401",
    year = "2006"
}

@article{Sharma:2021pkl,
    author = "Sharma, Atul",
    title = "{Twistor action for general relativity}",
    eprint = "2104.07031",
    archivePrefix = "arXiv",
    primaryClass = "hep-th",
    month = "4",
    year = "2021"
}

@article{Bittleston:2022nfr,
    author = "Bittleston, Roland and Skinner, David and Sharma, Atul",
    title = "{Quantizing the Non-linear Graviton}",
    eprint = "2208.12701",
    archivePrefix = "arXiv",
    primaryClass = "hep-th",
    doi = "10.1007/s00220-023-04828-0",
    journal = "Commun. Math. Phys.",
    volume = "403",
    number = "3",
    pages = "1543--1609",
    year = "2023"
}

@article{Sharma:2025ntb,
    author = "Sharma, Atul and Skinner, David",
    title = "{Chiral holography}",
    eprint = "2512.04152",
    archivePrefix = "arXiv",
    primaryClass = "hep-th",
    month = "12",
    year = "2025"
}

@article{Costello:2016mgj,
    author = "Costello, Kevin and Li, Si",
    title = "{Twisted supergravity and its quantization}",
    eprint = "1606.00365",
    archivePrefix = "arXiv",
    primaryClass = "hep-th",
    month = "6",
    year = "2016"
}

@article{Costello:2018zrm,
    author = "Costello, Kevin and Gaiotto, Davide",
    title = "{Twisted holography}",
    eprint = "1812.09257",
    archivePrefix = "arXiv",
    primaryClass = "hep-th",
    doi = "10.1007/JHEP01(2025)087",
    journal = "JHEP",
    volume = "01",
    pages = "087",
    year = "2025"
}

@article{Elliott:2020ecf,
    author = "Elliott, Chris and Safronov, Pavel and Williams, Brian R.",
    title = "{A taxonomy of twists of supersymmetric Yang{\textendash}Mills theory}",
    eprint = "2002.10517",
    archivePrefix = "arXiv",
    primaryClass = "math-ph",
    doi = "10.1007/s00029-022-00786-y",
    journal = "Selecta Math.",
    volume = "28",
    number = "4",
    pages = "73",
    year = "2022"
}

@article{Baulieu:1995bq,
    author = "Baulieu, Laurent",
    title = "{Field antifield duality, p form gauge fields and topological field theories}",
    eprint = "hep-th/9512026",
    archivePrefix = "arXiv",
    reportNumber = "PAR-LPTHE-95-55",
    month = "12",
    year = "1995"
}

@article{Bershadsky:1993cx,
    author = "Bershadsky, M. and Cecotti, S. and Ooguri, H. and Vafa, C.",
    title = "{Kodaira-Spencer theory of gravity and exact results for quantum string amplitudes}",
    eprint = "hep-th/9309140",
    archivePrefix = "arXiv",
    reportNumber = "HUTP-93-A025, RIMS-946, SISSA-142-93-EP",
    doi = "10.1007/BF02099774",
    journal = "Commun. Math. Phys.",
    volume = "165",
    pages = "311--428",
    year = "1994"
}

@article{Beem:2013sza,
    author = "Beem, Christopher and Lemos, Madalena and Liendo, Pedro and Peelaers, Wolfger and Rastelli, Leonardo and van Rees, Balt C.",
    title = "{Infinite Chiral Symmetry in Four Dimensions}",
    eprint = "1312.5344",
    archivePrefix = "arXiv",
    primaryClass = "hep-th",
    reportNumber = "YITP-SB-13-45, CERN-PH-TH-2013-311, HU-EP-13-78",
    doi = "10.1007/s00220-014-2272-x",
    journal = "Commun. Math. Phys.",
    volume = "336",
    number = "3",
    pages = "1359--1433",
    year = "2015"
}

@article{Baulieu:1990uv,
    author = "Baulieu, Laurent and Bellon, Marc P. and Ouvry, Stephane and Wallet, Jean-Christophe",
    title = "{Balatin-Vilkovisky analysis of supersymmetric systems}",
    reportNumber = "IPNO-TH-90-16, PAR-LPTHE-90-19",
    doi = "10.1016/0370-2693(90)90557-M",
    journal = "Phys. Lett. B",
    volume = "252",
    pages = "387--394",
    year = "1990"
}

@article{Wolf:2010av,
    author = "Wolf, Martin",
    title = "{A First Course on Twistors, Integrability and Gluon Scattering Amplitudes}",
    eprint = "1001.3871",
    archivePrefix = "arXiv",
    primaryClass = "hep-th",
    doi = "10.1088/1751-8113/43/39/393001",
    journal = "J. Phys. A",
    volume = "43",
    pages = "393001",
    year = "2010"
}

@article{Baulieu:2010ch,
    author = "Baulieu, Laurent",
    title = "{SU(5)-invariant decomposition of ten-dimensional Yang-Mills supersymmetry}",
    eprint = "1009.3893",
    archivePrefix = "arXiv",
    primaryClass = "hep-th",
    doi = "10.1016/j.physletb.2010.12.044",
    journal = "Phys. Lett. B",
    volume = "698",
    pages = "63--67",
    year = "2011"
}

@article{Ferber:1977qx,
    author = "Ferber, Alan",
    title = "{Supertwistors and Conformal Supersymmetry}",
    reportNumber = "EFI 77/50-CHICAGO",
    doi = "10.1016/0550-3213(78)90257-2",
    journal = "Nucl. Phys. B",
    volume = "132",
    pages = "55--64",
    year = "1978"
}

@article{Witten:2003nn,
    author = "Witten, Edward",
    title = "{Perturbative gauge theory as a string theory in twistor space}",
    eprint = "hep-th/0312171",
    archivePrefix = "arXiv",
    doi = "10.1007/s00220-004-1187-3",
    journal = "Commun. Math. Phys.",
    volume = "252",
    pages = "189--258",
    year = "2004"
}

@article{Boels:2006ir,
    author = "Boels, Rutger and Mason, L. J. and Skinner, David",
    title = "{Supersymmetric Gauge Theories in Twistor Space}",
    eprint = "hep-th/0604040",
    archivePrefix = "arXiv",
    doi = "10.1088/1126-6708/2007/02/014",
    journal = "JHEP",
    volume = "02",
    pages = "014",
    year = "2007"
}

@article{Witten:1988ze,
    author = "Witten, Edward",
    title = "{Topological Quantum Field Theory}",
    reportNumber = "IASSNS-HEP-87-72",
    doi = "10.1007/BF01223371",
    journal = "Commun. Math. Phys.",
    volume = "117",
    pages = "353",
    year = "1988"
}

@article{Costello:2013zra,
    author = "Costello, Kevin",
    title = "{Supersymmetric gauge theory and the Yangian}",
    eprint = "1303.2632",
    archivePrefix = "arXiv",
    primaryClass = "hep-th",
    month = "3",
    year = "2013"
}

@article{Gomis:1994he,
    author = "Gomis, Joaquim and Paris, Jordi and Samuel, Stuart",
    title = "{Antibracket, antifields and gauge theory quantization}",
    eprint = "hep-th/9412228",
    archivePrefix = "arXiv",
    reportNumber = "CCNY-HEP-94-03, KUL-TF-94-12, UB-ECM-PF-94-15, UTTG-11-94",
    doi = "10.1016/0370-1573(94)00112-G",
    journal = "Phys. Rept.",
    volume = "259",
    pages = "1--145",
    year = "1995"
}

@mastersthesis{Adamo:2013cra,
    author = "Adamo, Tim",
    title = "{Twistor actions for gauge theory and gravity}",
    eprint = "1308.2820",
    archivePrefix = "arXiv",
    primaryClass = "hep-th",
    type = "Other thesis",
    month = "8",
    year = "2013"
}

@article{Adamo:2017qyl,
    author = "Adamo, Tim",
    title = "{Lectures on twistor theory}",
    eprint = "1712.02196",
    archivePrefix = "arXiv",
    primaryClass = "hep-th",
    doi = "10.22323/1.323.0003",
    journal = "PoS",
    volume = "Modave2017",
    pages = "003",
    year = "2018"
}

@article{Woodhouse:1985id,
    author = "Woodhouse, N. M. J.",
    title = "{Real methods in twistor theory}",
    doi = "10.1088/0264-9381/2/3/006",
    journal = "Class. Quant. Grav.",
    volume = "2",
    pages = "257--291",
    year = "1985"
}

@article{Chalmers:1996rq,
    author = "Chalmers, G. and Siegel, W.",
    title = "{The Selfdual sector of QCD amplitudes}",
    eprint = "hep-th/9606061",
    archivePrefix = "arXiv",
    reportNumber = "ITP-SB-96-29",
    doi = "10.1103/PhysRevD.54.7628",
    journal = "Phys. Rev. D",
    volume = "54",
    pages = "7628--7633",
    year = "1996"
}

@article{Jarov:2025qhz,
    author = "Jarov, Seraphim",
    title = "{Twisted holography from the B-model on a 7-fold}",
    eprint = "2512.07412",
    archivePrefix = "arXiv",
    primaryClass = "hep-th",
    month = "12",
    year = "2025"
}

@article{boels_supersymmetric_2007,
    title = {Supersymmetric {Gauge} {Theories} in {Twistor} {Space}},
    volume = {2007},
    issn = {1029-8479},
    url = {http://arxiv.org/abs/hep-th/0604040},
    doi = {10.1088/1126-6708/2007/02/014},
    language = {en},
    number = {02},
    urldate = {2025-06-17},
    journal = {Journal of High Energy Physics},
    author = {Boels, Rutger and Mason, Lionel and Skinner, David},
    month = feb,
    year = {2007},
    note = {arXiv:hep-th/0604040},
    pages = {014--014},
}

\end{document}